\documentclass[prd,preprint,tightenlines,floatfix,showpacs,preprintnumbers,nofootinbib,eqsecnum]{revtex4}
 \usepackage[dvips,final]{graphicx}
  \usepackage{amssymb,amsmath,epsfig,bm,pifont}
   \usepackage{pstricks,pst-node,pst-text,pst-3d}
\usepackage{color}
 
  \newcommand{\BluTn}[1]{\textcolor{blue}{#1}}
   \newcommand{\RedTn}[1]{\textcolor{red}{#1}}
    \newrgbcolor{DarkGreen}{0.04 0.5 0.1}
\newcommand{\Ds}{\displaystyle}                           
 \renewcommand{\thesection}{\arabic{section}}

\newcommand{\be}{\begin{equation}}\newcommand{\ee}{\end{equation}}%
\newcommand{\bd}{\begin{displaymath}}\newcommand{\ed}{\end{displaymath}}
\newcommand{\bit}{\begin{itemize}}                        
 \newcommand{\eit}{\end{itemize}}                         
\newcommand{\ben}{\begin{enumerate}}                      
 \newcommand{\een}{\end{enumerate}}                       
\newcommand{\baa}{\begin{array}{lll}}                     
 \newcommand{\eaa}{\end{array}}                           
\newcommand{\ba}{\begin{eqnarray}}                        
 \newcommand{\ea}{\end{eqnarray}}                         
\newcommand{\la}{\label}                                  
\newcommand{\gev}[1]{\relax\ifmmode{\text{GeV}^{#1}}      
                     \else{GeV$^{#1}${ }}\fi}             
\newcommand{\Gev}{\relax\ifmmode{\text{GeV}}              
                     \else{GeV{ }}\fi}                    
\newcommand{\Mev}{\relax\ifmmode{\text{MeV}}              
                     \else{MeV{ }}\fi}                    
\def\MSbar{\relax\ifmmode\overline                        
            {\rm MS}\else{$\overline{\rm MS}${ }}\fi}     
\def\as{\relax\ifmmode \alpha_s\else{$ \alpha_s${ }}\fi}  
\def\abar{\relax\ifmmode{\bar{a}}\else{$\bar{a}${ }}\fi}  
\newcommand{\dRRNA}{\frac{d_F^{abcd}d_F^{abcd}}{N_A}}
\newcommand{\dRANA}{\frac{d_F^{abcd}d_A^{abcd}}{N_A}}
\newcommand{\dAANA}{\frac{d_A^{abcd}d_A^{abcd}}{N_A}}

\def\1{\hbox{{1}\kern-.25em\hbox{l}}}

\begin{document}
\thispagestyle{empty}
 \date{\today}
\preprint{\hbox{INR-TH-9/2014}}

\title{Generalization of the Brodsky-Lepage-Mackenzie optimization within 
the $\{\beta\}$-expansion and the Principle of Maximal Conformality}
 \author{A.~L. Kataev}
  \email{kataev@ms2.inr.ac.ru}
   \affiliation{Institute for Nuclear Research of the Academy
of Sciences of Russia, 117312, Moscow, Russia\\}

 \author{S.~V.~Mikhailov}
  \email{mikhs@theor.jinr.ru}
   \affiliation{Bogoliubov Laboratory of Theoretical Physics, JINR,
                141980 Dubna, Russia\\}
\begin{abstract}
We discuss generalizations of the BLM optimization procedure for renormalization group invariant quantities.
In this respect, we discuss in detail the features and construction of the
$\{\beta\}$--expansion representation instead of the standard perturbative series
with regards to the Adler $D$-function and Bjorken polarized sum rules obtained in order of 
${\cal O}(\alpha_s^4)$.
Based on the $\{\beta\}$--expansion we analyse different schemes of optimization,
including the corrected Principle of Maximal Conformality,
numerically illustrating their results.
We suggest our scheme for the series optimization and apply it to both the above quantities.
\end{abstract}
\maketitle

\section{Introduction}
\label{sec:intro}
The problem of  scale-scheme dependence ambiguities in  the renormalization group
(RG) calculations  \cite{Vladimirov:1979ib} remains important.
 In the past few years, a new extension of the
BLM scale-fixing approach \cite{Brodsky:1982gc},
called the Principle of Maximal
Conformality (PMC),  was started \cite{Brodsky:2011ig} and formulated
in more detail in \cite{Brodsky:2011ta,Mojaza:2012mf,Wu:2013ei,Brodsky:2013vpa}
with a variety of applications to phenomenologically oriented  studies.

Here we show that the PMC approach is
closely related  to the sequential BLM (seBLM) method, originally proposed  in \cite{MS04}
for the  analysis  of the N$^2$LO QCD prediction for the  quantities like $e^+e^-$-annihilation
R-ratio.
This method was based on the
renormalization group (RG) inspired representation of the $\{\beta\}$-expansion for perturbative
series,
the one later used for other purposes in \cite{Kataev:2010du,Kataev:2011zh}.
The  seBLM  was constructed as a generalization of the works
devoted to the extension of the BLM
${\rm \overline{MS}}$--type scale fixing prescription
to the level of  next-to-next-to-leading order (N$^2$LO) QCD corrections \cite{Grunberg:1991ac,Kataev:1992jm}
 and beyond \cite{Beneke:1994qe,Neubert:1994vb,Ball:1995ni,Hornbostel:2002af}.

In this paper, we will use the  $\{\beta\}$-expansion
representation and the seBLM method  to study the
$e^+e^-$-annihilation R-ratio, the related Adler function $D^\text{EM}$ of the
electromagnetic quark currents and the Bjorken sum rule $S^\text{Bjp}$
of the polarized lepton-nucleon deep-inelastic  scattering (DIS).
We will clarify  the concrete  theoretical shortcomings of
the PMC QCD  studies performed  in a number of works on the subject,
in particular, in Refs. \cite{Brodsky:2011ta,Mojaza:2012mf,Wu:2013ei,Brodsky:2013vpa},
and will present the results for the corrected PMC approach.

Certain problems of the misuse of the PMC approach to the Adler
function were  already emphasized  in \cite{Kataev:2013vua} but not recognized
in the recently published work \cite{Brodsky:2013vpa}.
We will clarify these theoretical problems in more detail and consider the existing
modification of the N$^2$LO  PMC analysis, 
based on application of the seBLM method, 
which allows one to reproduce the original
NLO BLM expression from the considerations performed in
\cite{Brodsky:2013vpa} and  already discussed
in  \cite{Kataev:2013vua}.
Note that the necessity of introducing modifications to the analysis of  \cite{Brodsky:2013vpa}
 starts to manifest itself from the level of taking into account the second order
perturbative corrections to the R-ratio evaluated analytically in
\cite{Chetyrkin:1979bj} in the  minimal subtractions (MS) scheme proposed in \cite{'tHooft:1973mm}.
This  result was also obtained numerically in \cite{Dine:1979qh} and confirmed
analytically in \cite{Celmaster:1979xr} by using  the ${\rm \overline{MS}}$-scheme of
\cite{Bardeen:1978yd}.
At the level of the third order corrections to $D^\text{EM}$,
analytically calculated in the ${\rm \overline{MS}}$ scheme
\cite{GorKatL,SurgSam} and confirmed  in the independent work \cite{Chetyrkin:1996ez},
there appear additional differences between the results of the PMC  and  the seBLM methods.

We present several  arguments in favour of
theoretical and phenomenological applications
of the form of the ${\beta}$-expanded
expressions for the RG invariant (RGI) quantities
 proposed in \cite{MS04} and applied in
\cite{Kataev:2010du}\footnote{Note,
that the $\{\beta\}$-expansion representation is related in part
to the considered in \cite{LovettTurner:1995ti} expansion of the perturbative
terms in the RG invariant (RGI) Green
functions through the powers of the first coefficient  of the
$\beta$-function}.
In this respect, let us mention the
  QCD generalization (in ${\rm \overline{MS}}$-scheme) of the
Crewther relation \cite{CR} based on the $\{\beta\}$-expansion \cite{Kataev:2010du}.
Using the results of these relations we obtain in a self-consistent way the
N$^2$LO $\{\beta\}$-expansion for $S^\text{Bjp}(Q^2)$
in QCD with $n_{\tilde{g}}$-numbers of  gluinos,
which can be checked by direct analytical calculations.

The article is organized as follows.
In Sec.~2, we define single-scale RG invariant quantities for the
$e^+e^-$-annihilation to hadrons and for the DIS inclusive  processes,
which will be studied in this work.
The existing theoretical relations between perturbative expressions for
these characteristics are also summarized.
In Sec.~3, the   $\{\beta\}$-expansion of the  RGI quantities,
proposed  in \cite{MS04} and  applied  in
\cite{Kataev:2010du,Kataev:2011zh}, is   reminded and discussed in detail.
Using the results of \cite{MS04} and the ``multiple power $\beta$-function''
QCD expression  \cite{Kataev:2010du}  for the ${\rm \overline{MS}}$-scheme  generalization of the
Crewther relation \cite{CR}  we provide the   arguments
 that this expansion is unique.
 The details of constructing the
$\{\beta\}$-expansions  for the Adler $D^\text{EM}$ function
 and for the $S^{Bjp}$  sum rule  are described
at the level of the  $O(a_s^3)$-corrections, where  $a_s=\alpha_s/(4\pi)$.
In Sec.~\ref{sec:Crewther.1}, we consider the relations between
certain terms of the $\{\beta\}$-expansion for $D^\text{EM}$
and $S^\text{Bjp}$, which will be  obtained from the Crewther
relation of ~\cite{CR} and its QCD generalization of
\cite{Kataev:2010du}, and present the concrete $\{\beta\}$-expanded
contributions to  the $D^\text{EM}$ function, R-ratio and
the $S^\text{Bjp}$ sum rule.

Using our  definition of the $\{\beta\}$-expansion representation
we correct the values of the  PMC coefficients and the scales in
the related powers of the PMC perturbative expressions  for the
Adler function $D^\text{EM}$ and  $R_{e^+e^-}$--ratio,
presented in \cite{Brodsky:2011ta,Mojaza:2012mf,Wu:2013ei,Brodsky:2013vpa},
and discuss their  correspondence
to  the results obtained in \cite{MS04,Kataev:2010du,Kataev:2013vua}.
The discussion of the results of the
BLM, seBLM,  PMC  procedures
together with the numerical estimates of the corresponding perturbation theory
(PT) coefficients and the couplings at new normalization scales are presented in
Sec.~\ref{sec:optBLM}.
It is demonstrated  that in spite of
its theoretical prominence following from  the
conformal symmetry relations even the corrected  PMC procedure does not improve
the convergence of perturbative series for the $R$-ratio
and for the $S^\text{Bjp}$ sum rule.
The methods of   further optimizations of these series,
which are  based on the $\{\beta\}$-expansion,  are elaborated in Sec. \ref{OptBLM}.
The technical results are presented in the appendices.

\section{Definitions of the basic Quantities}
Consider first the   Adler function $D^\text{EM}(Q^2)$, which is
expressed  through the two-point correlator of the electromagnetic vector currents
$j^{\rm EM}_{\mu} = \sum_i q_i\, \overline{\psi}_i\gamma_{\mu}\psi_i$ taken
at Euclidean $-q^2=Q^2$.
Here $q_i$ stands for the electric charge of the quark field $\psi_i$.
$D^{\rm EM}(Q^2)$
consists of  the sum of its nonsinglet (NS)  and singlet (S) parts
\begin{subequations}
\label{eq:D}
\ba
\label{DA}
\!\!\!\!D^{\rm EM}\left(\frac{Q^2}{\mu^2},a_s(\mu^2)\right)\!\! =\!\! \left(  \sum_i q_i^2
\right)\!\!
d_R D^{\rm NS}\left(\frac{Q^2}{\mu^2},a_s(\mu^2)\right)
\!+\! \left(\sum_i q_i \right)^2\!\! d_R D^{\rm S}\left(\frac{Q^2}{\mu^2},a_s(\mu^2)\right),
\ea
where
\ba
\label{DNS}
D^{\rm NS}\left(Q^2/\mu^2,a_s(\mu^2)\right)  & = &   1 + \sum_{l\geq 1} \ d^{\rm NS}_l\left(Q^2/\mu^2\right)
a_s^l(\mu^2)\,, \\
\label{DS}
~~D^{\rm S}\left(Q^2/\mu^2,a_s(\mu^2)\right) & = &  \frac{d^{abc}d^{abc}}{d_R} \sum_{l\geq 3} d_l^{\rm
S}\left(Q^2/\mu^2\right)
a_s^{l}(\mu^2).
\ea
\end{subequations}
Here $d_R$ is the dimension of quark representation of $SU(N_c)$ color gauge group
($d_R=N_c$) and $ d^{abc}=2Tr\left(\{\frac{\lambda_a}2,\frac{\lambda_b}2 \}\frac{\lambda_c}2\right)$
is the symmetric tensor.
Both the NS and S contributions to the Adler-function are the RGI quantities
calculable in the Euclidean domain.
After applying the RG equation
they can be represented as
\begin{subequations}
 \ba
\label{DNSRG}
D^{\rm NS}(Q^2/\mu^2,a_s(\mu^2)) &\stackrel{\mu^2= Q^2}{\longrightarrow} &D^{\rm NS}(a_s(Q^2))= 1 +
\sum_{l\geq 1} \ d^{\rm NS}_l a_s^l(Q^2) \\
~~D^{\rm S}(Q^2/\mu^2,a_s(\mu^2))   &\stackrel{\mu^2= Q^2}{\longrightarrow} &D^{\rm S}(a_s(Q^2))=
\frac{d^{abc}d^{abc}}{d_R} \sum_{l\geq 3} d_l^{\rm S}
a_s^{l}(Q^2).\label{DSRG}
\ea
 \end{subequations}
Due to  the cancellation
of the logarithms $\ln^k(Q^2/\mu^2 )$ with $k\geq l+1$ in the terms $ d^{a}_i\left(Q^2/\mu^2\right)$
(the superscript $a$ defines  the contributions to the  NS and S
parts of the $D^{\rm EM}$-function)
 the coefficients of $d_l^{a}\equiv d_l^{a}(1)$ are the  numbers in the
MS--like schemes.

 Let us emphasize that in this work we use
the perturbative expansion parameter
 $a_s(\mu^2)$
normalized as $a_s(\mu^2)\equiv \alpha_s(\mu^2)/4\pi$.
It  obeys
the RG equation with the consistently normalized  $SU(N_c)$-group
$\beta$-function
\be
\label{eq:beta}
 \mu^2\,\frac{\mathrm{d}}{\mathrm{d} \mu^2}
a_s(\mu) = \beta(a_s)=-a_s^{2} \sum_{i \ge 0} \beta_i a_s^{i}\, ,
\ee
where  $\beta_0 = (11/3C_\text{A} - (4/3) T_\text{R}n_f)$ while
other coefficients $\beta_i$ are presented in Appendix~\ref{App:A}.

The quantity related to the observable total cross-section of the
$e^+e\rightarrow{hadrons}$ process
$
 \Ds R_{e^+e^-}(s)
=
 \sigma(e^+e^-
 \to \mbox{hadrons})/\sigma(e^+e^- \to \mu^+\mu^-)
$
is measured in the Minkowski region ($s >0$); this can be obtained from
the $D^{\rm EM}$-function as
\begin{eqnarray}
 \label{eq:D->R}
   R_{e^+e^-}(s)&\equiv &   R(s,\mu^{2}=s)=
  \frac{1}{2\pi \textit{i}}
  \int_{-s-i\varepsilon}^{-s+i\varepsilon}\!
  \frac{D^{\rm EM}(\sigma/\mu^2;a_s(\mu^2))}
  {\sigma}\,
  d\sigma~
  \Bigg|_{\mu^2=s} =  \\ \nonumber
  &=& \left(  \sum_i q_i^2 \right) d_R \left (1+ \sum_{m\geq 1} r_m^{\rm NS} \, a_s^m(s)\right )+
  \left(  \sum_i q_i \right)^2 \frac{d^{abc}d^{abc}}{d_R} \sum_{n\geq 3} r_n^{\rm S} a_s^{n}(s)
\label{eq:FOPT}
\end{eqnarray}
The coefficients $r_m^{ a}$ for  the $a$ part ($a=$ NS or S) of $R_{e^+e^-}$ are associated with the
coefficients $d_l^{a}$ of the $D^{a}$ function
by the triangular matrix $T^{a}$  of the  relation
$r_m^{ a}=T_{ml}^{a} d_l^{ a}$, which will be discussed in subsection \ref{sec:$R$-ratio} and
Appendix \ref{App-D}.

The next  observable RGI quantity, we will be interested in,
is the   Bjorken polarized  sum rule $S^\text{Bjp}$. It is defined
by the integral over the difference of the spin-dependent structure functions of the polarized
lepton-proton
and lepton-neutron deep-inelastic scattering as
\be
S^\text{Bjp}(Q^2) =
\int_0^1 [g_1^{lp}(x,Q^2)-g_1^{ln}(x,Q^2)]dx
=\frac{g_A}{6}~C^\text{Bjp}(Q^2/\mu^2,a_s(\mu^2))~,
\label{gBSR}
\ee
where  $g_A$ is the nucleon axial charge as measured in the
neutron $\beta$-decay, and $C^\text{Bjp}(a_s)$ is coefficient function calculable within
perturbation
theory and not damped by the  inverse powers of $Q^2$, i.e., the leading twist term.

The application of the operator-product expansion (OPE)  method in the MS-like scheme
\cite{Gorishnii:1986gn}
 and the knowledge on the perturbative structure of the ${\rm \overline{MS}}$-scheme
QCD generalization of the quark-parton model Crewther relation \cite{CR} gained from articles
in \cite{BK93,GabadadzeKataev,Cr97,BKM03,Baikov:2010je,Baikov:2012zn,Kataev:2011im}
indicate  the existence of the previously undiscussed  singlet
contribution to $C^\text{Bjp}(a_s)$ \cite{Larin:2013}.
Using the results of this work we define the overall perturbative expression
for $C^\text{Bjp}$ as
\begin{equation}
C^\text{Bjp}(Q^2/\mu^2,a_s(\mu^2))=
C^\text{Bjp}_{\rm NS}(Q^2/\mu^2,a_s(\mu^2)) +
\left ( \sum_{i} q_i \right )C^\text{Bjp}_{\rm S}(Q^2/\mu^2,a_s(\mu^2))
\end{equation}
where the NS and S coefficient functions can be written down as
\ba
\label{CN}
C^\text{Bjp}_{\rm NS}(Q^2/\mu^2,a_s(\mu^2)) &=&    1 + \sum_{l\geq 1} \ c_l^{\rm NS}(Q^2/\mu^2)
a_s^l(\mu^2) \\
\label{CS}
C^\text{Bjp}_{\rm S}(Q^2/\mu^2,a_s(\mu^2))  &=&     \frac{d^{abc}d^{abc}}{d_R} \sum_{l\geq 3}
c_l^{\rm S}(Q^2/\mu^2)
a_s^{l}(\mu^2),
\ea
and have the following RG-improved form
\begin{subequations}
 \label{eq:CBj}
 \ba
\label{CBjpNS}
C^\text{Bjp}_{\rm NS}(Q^2/\mu^2,a_s(\mu^2)) &\stackrel{\mu^2= Q^2}{\longrightarrow}
&C^\text{Bjp}_{\rm NS}(a_s(Q^2))=
1 + \sum_{l\geq 1} \ c^{\rm NS}_l a_s^l(Q^2), \\
\label{CBjpS}
C^\text{Bjp}_{\rm S}(Q^2/\mu^2,a_s(\mu^2))   &\stackrel{\mu^2= Q^2}{\longrightarrow}
&C^\text{Bjp}_{\rm S}(a_s(Q^2))=
\frac{d^{abc}d^{abc}}{d_R} \sum_{l\geq 3} c_l^{\rm S}
a_s^{l+1}(Q^2), \\
C^\text{Bjp}(a_s(Q^2))&=& C^\text{Bjp}_{\rm NS}(a_s(Q^2))+C^\text{Bjp}_{\rm S}(a_s(Q^2)).
\ea
 \end{subequations}
The analytical expressions for the NLO and N$^2$LO  corrections to Eq.(\ref{CBjpNS})
in the ${\rm \overline{MS}}$-scheme
were  evaluated in \cite{Gorishnii:1985xm} and \cite{Larin:1991tj}, respectively,
while the corresponding  N$^3$LO ${\cal O}(a_s^4)$-correction
 was calculated in \cite{Baikov:2010je} (its direct analytical  form
 was also   presented  in \cite{Kataev:2010du}).
 The symbolic  expression for the  coefficient $c_4^{S}$
of  the ${\cal O}(a_s^4)$- correction to   the singlet contribution $C^\text{Bjp}_\text{S}(a_s)$
 of  the Bjorken polarized   sum rule was fixed  in \cite{Larin:2013} from the
 ${\rm \overline{MS}}$-scheme generalization of the Crewther relation,
 which will be presented below.

Let us also consider the Gross-Llewellyn Smith (GLS) sum rule of the
deep-inelastic neutrino-nucleon scattering.
Its leading twist perturbative QCD  expression can be defined as
\be
\label{GLS}
S_\text{GLS}(Q^2)=\frac{1}{2}\int_0^1[F_3^{\nu p}(x,Q^2)+F_3^{\nu n}(x,Q^2)]dx =
3C_\text{GLS}(Q^2/\mu^2,a_s(\mu^2))
\ee
where $F_3(x,Q^2)$ is  the structure functions of
the deep-inelastic neutrino-nucleon scattering process.
The coefficient function in the
RHS of Eq.(\ref{GLS}) also  contains both NS and S contributions, namely
\be
\label{GLSsr}
C_\text{GLS}(Q^2/\mu^2,a_s(\mu^2))
=C_\text{GLS}^{\rm NS}(Q^2/\mu^2,a_s(\mu^2))+C_\text{GLS}^{\rm S}(Q^2/\mu^2,a_s(\mu^2))~~~.
\ee
As a consequence of the chiral invariance, which can be restored in the
dimensional regularization \cite{'tHooft:1972fi}
by means of additional finite renormalizations 
(for their consequent evaluation
in high-loop orders see, e.g.,
 \cite{Gorishnii:1985xm,Larin:1991tj,Antoniadis:1979us,Trueman:1979en}),
the NS contributions to the leading twist coefficient function of $S_\text{GLS}(Q^2)$
coincide with a similar NS perturbative  contribution $S^\text{Bjp}(Q^2)$, namely
\ba
\nonumber
C_\text{GLS}^{\rm NS}(Q^2/\mu^2,a_s(\mu^2))&\equiv&
C^\text{Bjp}_{\rm NS}(Q^2/\mu^2,a_s(\mu^2)) =
1 + \sum_{l\geq 1} \ c_l^{\rm NS}(Q^2/\mu^2) a_s^l(\mu^2) \\ \label{Cns}
&\stackrel{\mu^2= Q^2}{\longrightarrow}&C^\text{Bjp}_{\rm NS}(a_s(Q^2))= 1 + \sum_{l\geq 1} \ c^{\rm
NS}_l a_s^l(Q^2)~.
\ea
The fulfilment of this identity was explicitly demonstrated in the existing
analytical NLO and N$^2$LO calculations
in \cite{Gorishnii:1985xm,Larin:1991tj} and used
as the input in the process of determination of the analytical
expression for the ${\cal O}(a_s^4)$ corrections to $S_\text{GLS}(Q^2)$ \cite{Baikov:2012zn}.

The second (singlet-type)  contribution to the coefficient function of
Eq.(\ref{GLSsr}) has the following form:
\ba
\nonumber
C_\text{GLS}^{\rm S}(Q^2/\mu^2,a_s(\mu^2))&=&n_f\frac{d^{abc}d^{abc}}{d_R}\sum_{l\geq 3}
\bar{c_l}^(Q^2/\mu^2)a_s^{l}(\mu^2) \\
\label{GLSSI}
&\stackrel{\mu^2= Q^2}{\longrightarrow}&C_\text{GLS}^{\rm
S}(a_s(Q^2))=n_f\frac{d^{abc}d^{abc}}{d_R}\sum_{l\geq 3} \bar{c_l}
a_s^{l}(Q^2)
\ea
where $\bar{c}_3$ and $\bar{c}_4$ were evaluated analytically
in \cite{Larin:1991tj} and \cite{Baikov:2012zn}, respectively.

The application of the OPE approach to the
three-point functions of axial-vector-vector currents
(see \cite{GabadadzeKataev,Cr97,BKM03,Kataev:2011im,Larin:2013})
leads to  the following ${\rm \overline{MS}}$-scheme  QCD generalization
of the Crewther relation (CR) between the introduced above different
coefficient functions of the annihilation and deep-inelastic
scattering processes:
\begin{subequations}
 \label{eq:Crether}
\ba
\label{identity}
C^\text{Bjp}(a_s)D^{\rm NS}(a_s)&\equiv& C_\text{GLS}(a_s)[D^{\rm NS}(a_s)+n_fD^{\rm S}(a_s)] \\
\label{factorize}
&=&  \1 + \frac{\beta(a_s)}{a_s}\,\cdot K(a_s),
\ea
\end{subequations}
where $\1$ was derived in \cite{CR} using
the conformal symmetry,
$\beta(a_s)$ is the RG $\beta$-funtion, $a_s=a_s(Q^2)$ and the polynomial $K$ is
\be
K(a_s) =
a_s\,K_1 +a_s^2\,K_2 +a_s^3\,K_3+ O(a_s^4)~.
\ee
It contains the coefficients  $K_1$ and $K_2$, obtained in \cite{BK93},
while the analytical expression for the coefficient $K_3=K_3^{\rm NS}+K_3^{\rm S}$
is the sum of  the NS- and S-terms, which are given in
\cite{Baikov:2010je} and \cite{Baikov:2012zn}, respectively.
Note that Eq.(\ref{identity}) was first published in
\cite{Kataev:2011im} without taking into account singlet-type
contributions to $C^\text{Bjp}$.
Their more careful analysis of \cite{Larin:2013} fixes  the $\beta_0$-dependent analytical
expression of the ${\cal O}(a_s^4)$ contribution to $C^\text{Bjp}_{\rm S}$ \footnote{In QED
the validity of Eq.(\ref{identity}) follows from the considerations of
Ref.~\cite{ACGJ}.}.

The result of \cite{Larin:2013} and
the  general Eq.(\ref{identity})
is  not yet  confirmed by direct analytical calculations.
In our further studies we will use the product of their NS parts and related 
to this product results of the expansion in Eq.(\ref{factorize}), 
reformulated in \cite{Kataev:2010du}.

\section{General $\beta$-expansion structure of observables}
\label{sec:expantion structure}
\subsection{Formulation of the approach}
 \label{subsec:Formulation}

To clarify the main ideas of the $\{\beta\}$-expansion representation
proposed in \cite{MS04}
for the perturbative coefficients of the
RGI quantities, let us consider  the NS part of the
Adler function, $D^\text{NS}$.
Its  expression can be rewritten as
$D^{\rm NS}=1+d^\text{NS}_1 \cdot \sum_{n\geq 1}d_la_s^{l}$,
where
$d^\text{NS}_1=3{\rm C_F}$ is the overall normalization factor.
Within the $\{\beta\}$-expansion approach
the coefficients $d_n$, originally fixed in the ${\rm \overline{MS}}$-scheme,
are expressed as
\begin{subequations}
\label{eq:d_beta}
\begin{eqnarray}
\label{eq:d_1}
d_1&=&d_1[0]=1\, , \\
d_2&=&\! \beta_0\,d_2[1]
  + d_2[0]\, ,\label{eq:d_2}\\
  d_3
&=&\!
  \beta_0^2\,d_3[2]
  + \beta_1\,d_3[0,1]
  +       \beta_0 \,  d_3[1]
  + d_3[0]\, ,\label{eq:d_3} \\
  d_4
   &=&\! \beta_0^3\, d_4[3]
     + \beta_1\,\beta_0\,d_4[1,1]
     + \beta_2\, d_4[0,0,1]
     + \beta_0^2\,d_4[2]
     + \beta_1  d_4[0,1]
     + \beta_0\,d_4[1] \nonumber \\
   && \phantom{\beta_0^3\, d_4[3]+ \beta_1\,\beta_0\,d_4[1,1]+ \beta_0^2\,}~+d_4[0]\,,
       \label{eq:d_4} \\
   &\vdots& \nonumber \\
 d_{N}
   &=&\! \! \! \! \! \!~~\beta_0^{N-1}\! d_{N}[N\!-\!1]+ \cdots + d_N[0]\,,
\label{eq:d_n}
\end{eqnarray}
\end{subequations}
where the first  argument of the expansion  elements $d_n[n_0,n_1,\ldots]$ indicates its
multiplication to
the $n_0$-th power of the first coefficient $\beta_0$ of the
RG $\beta$-function, namely to the $\beta_0^{n_0}$ term.
The second argument $n_1$ determines the power
of the second  multiplication factor,
namely $\beta_1^{n_1}$, and so on.
The elements
$d_n[0]\equiv d_n[0,0,\ldots,0]$
define  ``refined'' $\beta_i$-independent
corrections
with  powers $n_i=0$ of  all their $\beta_i^{n_i}$ multipliers.
These  elements  coincide with  expressions for the
coefficients  $d_n$ in the  imaginary
situation of the nullified QCD   $\beta$--function in
all orders of perturbation theory.
This case  corresponds to the effective  restoration of the conformal symmetry limit
of the bare $SU(N_c)$ model  in the case when all normalizations are not
considered.
This limit,
extensively discussed in \cite{Kataev:2013vua},  will be considered here as a technical trick.
The origins of other elements in Eqs. (\ref{eq:d_beta}) were considered in \cite{MS04}.

The first elements $d_{i}[i\!-\!1]$  of the expansions of Eqs.(\ref{eq:d_2}-\ref{eq:d_n})
arise from the diagrams with a maximum number of the ``fermion 1-loop  bubble'' insertions
and applications of the Naive Non-Abelianization (NNA) approximation \cite{BroGro95}).
In the case of $D^{\rm NS}$-function they can be
obtained from the result in \cite{BK93}, which follow from renormalon-type
calculations in \cite{Broadhurst:1992si,Beneke:1992ch}.

It would be stressed that the terms   $\beta_0\, d_3[1]$,
 $\beta_1 d_4[0,1]$, $\beta_0  d_4[1]$
 in Eqs.(\ref{eq:d_beta}) were not taken into account in the
variant of the $\{\beta\}$-expansion method used
in \cite{Brodsky:2011ta,Mojaza:2012mf,Wu:2013ei}.
The omitting of these terms leads to the results, which
should be corrected by including these terms in the self-consistent variant of the PMC analysis.

In high order of perturbation theory one should also consider a similar  expression   of  the singlet
part
$D^{\rm S}=d^{\rm S}_3 \cdot \sum_{j\geq 3}\bar{d}_j a_s^{(j)} $ with  the
normalization factor $d^{\rm S}_3= 11/3 -8\zeta_3$ evaluated first in the
QED work \cite{Gorishnii:1990kd},
and the related normalizations of the defined  coefficients in Eq.(\ref{GLSSI}),
namely  $\bar{d}_j=d_j^{\rm S}/d^{\rm S}_3$.
The $\{\beta\}$-expanded coefficients of this RGI-invariant quantity are  expressed as
\begin{subequations}
 \begin{eqnarray}
\bar{d}_3 &=& \bar{d}_3[0] =1  \label{eq:d31}  \\
\bar{d}_4&=&\beta_0\,\bar{d}_4[1]  + \bar{d}_4[0]\,\label{eq:dS_4} ,\\
  &\vdots&   \nonumber \\
\bar{d}_{j+2}
   &=&\beta_0^{j-1}\! \bar{d}_{j+2}[j-1]+ \cdots + \bar{d}_{j+2}[0] . \label{eq:dS_n}
\end{eqnarray}
 \end{subequations}

The same ordering in the $\beta$-function coefficients can be applied to the
coefficients $c_n$ for the NS coefficient function of the deep-inelastic
sum rules   $C^{\rm NS}$ of Eq.(\ref{Cns}) and to
the singlet contribution  $C^{\rm S}$ to the GLS sum rule (see Eq.~(\ref{GLSSI}).
Moreover, it is possible to show that the elements of the corresponding $\{\beta\}$-expansions
$d_n[n_0,n_1,\dots]$ and $c_n[n_0,n_1,\dots]$ are closely related \cite{Kataev:2010du}.
We will return to a more detailed discussion of this property a bit later.

The above $\{\beta\}$--expansion can be interpreted as a ``matrix'' representation for
the RGI quantities:
For the quantity $D^\text{NS}$ expanded up to an order of $N$, $D^\text{NS}= \sum_{n=1}^{N} a_s^n\sum_{i\geq 0}
D_{ni}^\text{NS} B^{(i)}$
which is  related to the traditional ``vector'' representation, $D^\text{NS}= \sum_{n=1}^{N} a_s^n d_n$
with  $d_{n}=\sum_{i} D_{ni}^\text{NS} B^{(i)}$.
Here $B^{(i)}$ are the elements that express the structure of
$\{\beta\}$-expanded  perturbative coefficients and are
convolved with the matrix elements $D_{ni}^\text{NS}=d_n[\ldots]$.
In the case of consideration of the ``refined'' $\beta_i$-independent
corrections  $d_n[0]\equiv D_{n0}$ and  $B^{(0)}=1$.
 The similar matrix representation  can be written down for the singlet part $D^\text{S}$ with
the $\{\beta\}$-expanded coefficient defined in Eqs.(\ref{eq:d31}-\ref{eq:dS_n}).

Note that  the matrix representation contains new dynamical information
about the RGI invariant quantities, which is not contained in the vector one.
Thus, Eqs.~(\ref{eq:d_2}),(\ref{eq:dS_4}) can
be considered as  the initial points to apply the standard BLM procedure.
The generalization of the BLM procedure to higher orders can be
constructed using  the $\{\beta\}$--expansions of higher order coefficients
of Eqs. (\ref{eq:d_beta}) \cite{MS04}.
However, starting with the N$^2$LO the explicit  solution of this problem is nontrivial.

\subsection{Explicit  determination of the structures  of the
$\{\beta\}$-expanded series for $D^\text{NS}$}
\label{subsec:Explicit D}

Let us start the discussion of application of the $\{\beta\}$-expansion procedure in the NLO.
Imagine that we deal with the perturbative quenched QCD (pqQCD)  approximation for
the $D^{\rm NS}$-function in the NLO.
It is described
by the contributions of the three-loop photon vacuum polarization
diagrams with closed   external loop, formed by quark-antiquark
pair and  connected by internal gluon propagators, which do not
contain any internal quark-loop  insertions.
In this theoretical approximation the coefficient $d_2$ takes the following form:
\be \label{eq:d_2pqQCD}
d_2 \to d_2^{pqQCD}=-\frac{\rm C_F}2+ \left(\frac{123}2-44\zeta_3\right)\frac{\rm C_A}{3}
~\mbox{with}~\beta_0=\frac{11}3C_{\rm A}.
\ee
In  this case, it is unclear how to perform  the standard BLM
scale-fixing prescription in the NLO approximation.
Indeed, it is not clear what is the expression for the $d_2[1]$ coefficient of the
$\beta_0$-term of Eq.(\ref{eq:d_2}) in the expression for $d_2^{pqQCD}$.
To obtain explicitly the elements of the expansion (\ref{eq:d_2})
and extract the $\beta_0$-term in (\ref{eq:d_2pqQCD}),
one should take into account the quark-antiquark one-loop insertion in internal gluon
lines of the three-loop approximation for the  hadronic vacuum polarization function.
This is equivalent to taking into account in the pqQCD model 
of the interacting with gluons of internal 
quark loops with $\bm{n_f}$ number of active quarks.
The corresponding parameter $\bm{n_f}$ can be considered as  \textit{a mark}
 of the charge renormalization by the  quark-antiquark pair.
It  enters into  both $d_2$ and $\beta_0$ expressions  and allows one to extract
\textit{unambiguously} the expression for  $d_2$ proportional to the $\beta_0$-term in the
${\rm \overline{MS}}$-scheme.
Indeed, fixing $T_R=T_F=\frac1{2}$ we obtain
\be \label{eq:d_2qua}
d_2 =-\frac{\rm C_F}2 + \left(\frac{11\cdot 11+2}2-44\zeta_3\right)\frac{\rm C_A}{3}-
\left(\frac{11}2-4\zeta_3\right)\frac{2}{3}\bm{n_f}~\mbox{with}
~\beta_0=\frac{11}3C_{\rm A}-\frac2{3} \bm{n_f}.
\ee
To get the appropriate
expression of the coefficient $d_2$ one should take into account
the  1-loop renormalization of charge. As the result,
we immediately obtain from Eq.~(\ref{eq:d_2qua}) the following
expression for the coefficient of Eq.(\ref{eq:d_2}):
\begin{subequations}
\be
\label{eqd2}
d_2 =\frac{\rm C_A}3-\frac{\rm C_F}2+ \left(\frac{11}2-4\zeta_3\right)
\left(\frac{11}{3}{\rm C_A}-\frac{2}{3}\bm{n_f} \right)
\ee
where
\be
\label{terms}
d_2[0]= \frac{\rm C_A}3-\frac{\rm C_F}2,~~~~d_2[1]=\frac{11}2-4\zeta_3~~~.
\ee
 \end{subequations}
This decomposition  corresponds to the case of the  standard BLM
consideration in the ${\rm \overline{MS}}$-scheme \cite{Brodsky:1982gc}.
Note that for  $\bm{n_f}=0$   this decomposition  remains valid for
the case of pqQCD  (QCD at $\bm{n_f}=0$) and  leads to Eq.(\ref{terms}).

Any additional modifications of QCD, say, by means of  introducing
into considerations   $\bm{n_{\tilde{g}}}$ multiplets
of strong interacting gluino (the element of the MSSM model),
will change in the NLO expression for the  considered RGI invariant quantity
the content  of the $\beta_0$-coefficient in the expression for $d_2$,
calculated in the ${\rm \overline{MS}}$-scheme,
but not the ``refined'' element and  the coefficients at $\beta_0$ of Eq.(\ref{eq:d_2pqQCD}).
Using the result  $\beta_0$  for the $\beta$-function
with the $\bm{n_{\tilde{g}}}$ multiplet of strong interacting  gluino
(see Eq.~(\ref{eq:beta-b0})) and the $D^{\rm NS}$-function in the same
model ( presented
in  Eq.(\ref{eq:D-2}) of the  Appendix~\ref{App:A}),
the same result (\ref{terms}) for decomposition can unambiguously be obtained using
the additional  \textit{marks} in Eq.(\ref{eqd2}), namely the number of
strong-interacting gluino  $\bm{n_{\tilde{g}}}$.
Indeed, combining the result
\begin{subequations}
 \label{eq:2nf-ng}
  \ba
 \label{eq:d2nf-ng}
d_2 =\frac{\rm C_A}3-\frac{\rm C_F}2+ \left(\frac{11}2-4\zeta_3\right)
\left(\frac{11}{3}{\rm C_A}-\frac{2}{3}\bm{n_f} \right)
-\left(11-8\zeta_3\right) \bm{n_{\tilde{g}}}\frac{C_A}{3}  \,,
\ea
with
\be
\label{beta:ngnf}
\beta_0\left(\bm{n_f}, \bm{n_{\tilde{g}}}\right)=
    \frac{11}{3} C_A - \frac{2}{3}\left(\bm{n_f} + \bm{n_{\tilde{g}}} C_A\right)
\ee
 \end{subequations}
we get the expressions for $d_2[0]$ and $d_2[1]$, which are
identical to the ones presented in Eq.(\ref{terms}).
Note that these results can be obtained from Eq.(\ref{eq:d2nf-ng}) and Eq.(\ref{beta:ngnf})
with gluino degrees of freedom only ($\bm{n_{\tilde{g}}} \neq 0,~\bm{n_f}=0$),
or only  with the quark ones ($\bm{n_f}\neq 0,~\bm{n_{\tilde{g}}}=0$),
or with taking into account both of them.
The reason of this unambiguity is that the interaction of any new
particle accumulated here in the charge renormalization
is determined by the universal gauge group $SU(N_c)$.

All these possibilities  give us a simple tool to
restore the $\beta_0$-term in the NLO following the BLM precription \cite{Brodsky:1982gc}.
Thus, in the NLO we may switch off the gluino degrees of freedom.
However, to get the  $\{\beta\}$ expansion of the N$^2$LO term in the form of
Eq.~(\ref{eq:d_3}) we cannot use the quark degrees of freedom only.
Indeed, in this case, we face a problem similar to that which arises in the process of
$\{\beta\}$ decomposition of  the pqQCD expression for  $d_2^{pqQCD}$ in Eq.~(\ref{eq:d_2pqQCD})
discussed above.

The $\{\beta\}$--expanded  form  for the   $d_3$-term  was obtained in
Ref. \cite{MS04} by means of a careful consideration
of the analytical ${\cal O}(a_s^3)$ ${\rm \overline{MS}}$-scheme expression for the
Adler function $D^{\rm NS}(a_s,\bm{n_f}, \bm{n_{\tilde{g}}})$
with the $\bm{n_{\tilde{g}}}$ QCD interacting MSSM gluino multiplets
obtained in  \cite{Chetyrkin:1996ez} and presented in
Eq. (\ref{eq:D1-3})\footnote{The evaluated in \cite{Chetyrkin:1996ez} N$^2$LO analytical
result for the   gluino contribution was confirmed in \cite{Clavelli:1996zm}.}
together with the corresponding two-loop  $\beta$-function, $\beta(\bm{n_f}, \bm{n_{\tilde{g}}})$,
see~Eqs.~(\ref{eq:beta-b0},~\ref{eq:beta-b1}).

Let us consider this procedure  in more detail.
The element $d_3[2]$,  which is proportional to the maximum power $\beta_0^2$ in (\ref{eq:d_3}),
can be fixed in a straightforward way, using the results in \cite{BK93}.
Then one should separate the contributions of $\beta_1\, d_3[0,1]$ and of
$\beta_0\,d_3[1]$ to the $d_3$-term.
They  both are linear in the  number of quark  flavours $\bm{n_f}$,
therefore, they could not be disentangled directly.
Their separation is possible if one takes into account additional degrees of freedom,
\textit{e.g.}, the gluino contributions mentioned above
for both the quantities (the additional \textit{mark} appears),
namely for the  $D^{\rm NS}$-function from Eqs.(\ref{D3:gl}) and for the first two coefficients
of the $\beta$-function from Eqs.(\ref{eq:beta-b0},\ref{eq:beta-b1}).
In  this way, using two equations one can get the explicit form for the functions
$\bm{n_f}= n_f(\beta_0,\beta_1)$ and $\bm{n_{\tilde{g}}}=
 n_{\tilde{g}}(\beta_0,\beta_1)$.
Finally, substituting these functions in
$D=D(a_s, n_f(\beta_0,\beta_1),n_{\tilde{g}}(\beta_0,\beta_1))$
its  $\{\beta\}$-expanded expression was ontained in
\cite{MS04},
\begin{subequations}
 \begin{eqnarray}
\nonumber
D^\text{NS}(a_s,n_f, n_{\tilde{g}})
&=&1+  a_s(3C_F) + a_s^2(3C_F)\cdot
          \left\{\frac{\rm C_A}3-\frac{\rm C_F}2+
          \left(\frac{11}2-4\zeta_3\right)\beta_0\left(n_f, n_{\tilde{g}}\right) \right\}\\
&&+a_s^3(3C_F)\cdot
          \left\{ \left(\frac{302}9-\frac{76}3\zeta_3\right)\beta_0^2\left(n_f,
          n_{\tilde{g}}\right)+
          \left(\frac{101}{12}-8\zeta_3\right)\beta_1\left(n_f, n_{\tilde{g}}\right)\right. \nonumber
          \\
&&        \left.  +\left[{\rm C_A}\left(\frac{3}4 + \frac{80}3\zeta_3 -\frac{40}3\zeta_5\right) -
    {\rm C_F}\left(18 + 52\zeta_3 - 80\zeta_5\right)\right]\beta_0\left(n_f,
    n_{\tilde{g}}\right)\right. \nonumber \\ \label{d3-beta}
&&     \left. + \left(\frac{523}{36}-
        72 \zeta_3\right){\rm C_A^2}
    +\frac{71}3 {\rm C_A C_F} - \frac{23}{2} {\rm C_F^2}\right\}~,
\end{eqnarray}
with (see Eqs.(\ref{eq:beta0-3}) in Appendix\ref{App:A})
\begin{equation}
  \beta_1\left(n_f, n_{\tilde{g}}\right)
     = \frac{34}{3}C_A^2
       - \frac{20}{3}C_A \left( T_R n_f + \frac{n_{\tilde{g}} C_A}{2}\right)
      -4\left( T_R n_f C_\text{F}+ \frac{n_{\tilde{g}} C_A}{2} C_A\right).
\end{equation}
  \end{subequations}
Note that in order to write down the ${\cal O}(a_s^4)$ coefficient of $D^{\rm NS}$,
analytically evaluated in the ${\rm \overline{MS}}$-scheme in
\cite{Baikov:2010je} for the case of $SU(N_c)$
in a similar $\{\beta\}$-expanded form of Eq.(\ref{eq:d_4}),
it is necessary to perform additional calculations,
which generalize this result to the case of
 $SU(N_c)$ with $n_{\tilde{g}}$ multiplets of  gluino.
Then, one should combine this possible (but not yet existing)
generalization  with  already available
analytical expression for   the $\beta_2(n_f,n_{\tilde{g}})$-coefficient
of the $\beta$-function in this model, analytically obtained in
the ${\rm \overline{MS}}$-scheme in \cite{Clavelli:1996pz}.

\subsection{Does $\{\beta\}$-expansion have any ambiguities ?}
 It is instructive to discuss here an attempt \cite{Mojaza:2012mf,Brodsky:2013vpa} to obtain the
 elements
 $d_n[\ldots ]$ in a different way.
 This is based on the expression for $D$, rewritten in \cite{Baikov:2012zm} for the usability of
 current 5-loop
 computation in the form
 \ba
 D^\text{EM}(a_s) =  12\pi^2\, \left(\gamma^\text{EM}_{\rm ph}(a_s)
-
 \beta( a_s) \frac{d }{d a_s}\Pi^\text{EM}(a_s)
\right).
\label{Adler:master:eq}
 \ea
Here $\Pi^\text{EM}(a_s)=\Pi^\text{EM}(L, a_s)\equiv d_R/(4\pi)^2 \sum_{i\geq 0} \Pi_i a_s^i$ is the
polarization function of electromagnetic currents at
$L\equiv \ln(Q^2/\mu^2)=0$, and $\gamma^\text{EM}_{\rm ph}\equiv 1/(4\pi)^2 \sum_{j\geq 0} \gamma_j a_s^j$ is the anomalous dimension of the photon field.
In our notation Eq.~(\ref{Adler:master:eq}) leads to the expansion for $D^\text{NS}$,
\ba
D^\text{NS}(a_s)=
1+ 3C_\text{F} a_s + \left(12\gamma_2 +3\beta_0 \Pi_1 \right)a_s^2 +
\left(48\gamma_3 +3\beta_1 \Pi_1+24\beta_0 \Pi_2 \right)a_s^3+\ldots,
\label{Adler:master:exp}
\ea
where the ingredients of the expansion, $\gamma_i,~\Pi_j$, were calculated
in \cite{Baikov:2012zm} up to $i=4,~j=3$,
 and  we take corresponding NS projection in the RHS of Eq.(\ref{Adler:master:eq}).
The renormalization of the charge certainly contributes to the 3-loop anomalous
dimension $\gamma_2$; therefore, it contains a $\beta_0$-term also
(one can make sure from the inspection  of the explicit formula for $\gamma_2$
in Eq.~(3.12) in \cite{Baikov:2012zm} and even  in Eq.(10) in \cite{Dine:1979qh}).
Taking into account the explicit form of $\gamma_2$ and $\Pi_1$ in (\ref{Adler:master:exp})
one can recalculate  the well-known
decomposition for $D^\text{NS}$ in order ${\cal O}(a_s^2)$
\ba
D^\text{NS}(a_s)=
1+ 3C_\text{F}\cdot a_s + 3C_\text{F}\cdot \left(\beta_0 d_2[1]+ d_2[0]\right)a_s^2 + O(a_s^3)
\ea
in full accordance with the result in \cite{Brodsky:1982gc} and Eq.~(\ref{eq:d_2}) (for the related
discussions see Ref.
\cite{Kataev:2013vua} as well).

Unfortunately the authors of \cite{Brodsky:2013vpa} claim,
basing on a formal correspondence,
that the coefficient of $\beta_0$ is only the term $\Pi_1/C_\text{F}$ in
Eq.~(\ref{Adler:master:exp})
(with the above notation at $d_1$ normalized by unity),
while the ``conformal term'' is $4 \gamma_2/C_\text{F}$ 
(see Eq.~(48a-b) in \cite{Brodsky:1982gc}),
 which in reality is not true.
The comparison of these terms
\ba
d_2[1]=\frac{11}2-4\zeta_3\approx 0.69177&~ \Leftrightarrow ~&
\Pi_1/C_\text{F}=\RedTn{\frac{55}{12}-4\zeta_3\approx - 0.22489}\,; \label{eq:d2-comp}\\
\label{gamma2}
d_2[0]= d_2[0]= \frac{\rm C_A}3-\frac{\rm C_F}2&~
\Leftrightarrow ~&4\gamma_2/C_\text{F}=\frac{11}{12}\beta_0-\frac{\rm C_F}2
\ea
shows that they differ even in sign in (\ref{eq:d2-comp}) 
[compare $\Pi_1/C_\text{F}$ with $d_2[1]$].
The considerations of the lead to a shift of the BLM scale $Q^2_\text{BLM}$ 
in the opposite direction, $Q^2_\text{BLM} \geq Q^2$, 
in comparison with the standard value
$Q^2_\text{BLM}=\exp\left(-d_2[1]\right) Q^2\approx Q^2/2 $
(see the discussion after Eq.~(\ref{1-stage-n}) in Sec.\ref{sec:optBLM}). Moreover,
we demonstrate  in  Eq.(\ref{gamma2}) that $\gamma_2$ is not ``conformal'' and depends on $\beta_0$.

\section{partial $\beta$-expansion elements for $D,~C$, and $R$}
Here we extend our knowledge about the $\beta$-expansion elements on NS part of the Bjorken
$C^\text{Bjp}$ basing on CR Eq.(\ref{factorize}) for $D^\text{NS}$ and  $C^\text{Bjp}_\text{NS}$.
\subsection{What constraints Crewther relation gives}
\label{sec:Crewther.1}
In the case \textbf{of} the $\beta$-function has  identically zero coefficients
$\beta_i=0$ for $i\geq 0$ the generalized CR (\ref{factorize}) returns to its initial form
 \cite{CR}
\begin{equation}
\label{NC}
 D^\text{NS}_0 \cdot C^\text{Bjp}_0 = \1,
\end{equation}
where the expansions for the functions $D^\text{NS}_0$ and $C^\text{Bjp}_0$,
analogous to the ones of Eq.~(\ref{eq:d_2}-\ref{eq:d_n}),
contain the coefficients of  genuine content only,
namely,  $d_n~(c_n) \equiv d_n[0]~(c_n[0])$.
Equation (\ref{NC}) provides an evident relation between the genuine  elements in any loops,
namely,
\begin{equation}
c^\text{NS}_n[0] + d^\text{NS}_n[0] + \sum_{l=1}^{n-1} d^\text{NS}_l[0] c^\text{NS}_{n-l}[0]=0,
\label{eq:CI-PT0}
\end{equation}
where $d^\text{NS}_n[0]=d^\text{NS}_1 \cdot d_n[0]$ and  $c^\text{NS}_n[0]=c^\text{NS}_1\cdot
c_n[0]$
in virtue of the normalization condition.
From Eq.~(\ref{eq:CI-PT0}) at $n=1$ immediately follows that 
$c^\text{NS}_1=-d^\text{NS}_1$.
The relation (\ref{eq:CI-PT0}) can be used
to obtain the unknown genuine parts  of the
4-loop term   $c^\text{NS}_3[0]$,
through the 4-loop results already known from the analysis in \cite{MS04}:
\begin{subequations}
 \label{eq:c3}
\ba
\label{eq:NSc3-NSd3}
c^\text{NS}_3[0] &=& - d^\text{NS}_3[0]+2d^\text{NS}_1 d^\text{NS}_2[0]-(d^\text{NS}_1)^3, \\
 && \text{or, in the other normalized terms,} \nonumber \\
c_3[0] &=&  d_3[0]-2d^\text{NS}_1 d_2[0]+(d^\text{NS}_1)^2, \label{eq:c3-d3}
\ea
 \end{subequations}
It is useful to relate the unknown elements $c^\text{NS}_4[0],d^\text{NS}_4[0]$ in a 5-loop
calculation with the known elements
of the 4-loop results, viz,
\begin{equation}
\label{eq:c4-d4}
c^\text{NS}_4[0] + d^\text{NS}_4[0]=
2d^\text{NS}_1
d^\text{NS}_3[0]-3(d^\text{NS}_1)^2d^\text{NS}_2[0]+(d^\text{NS}_2[0])^2+(d^\text{NS}_1)^4.
\end{equation}

Let us consider now the generalized CR in Eq.(\ref{factorize}), which includes the terms
proportional to the conformal anomaly,
$\beta(a_s)/a_s$, appearing due to violation of the
the conformal symmetry in the renormalized ${\rm SU(N_c)}$ interaction (in the ${\rm
\overline{MS}}$-scheme).
 As was shown in \cite{Kataev:2010du}, this relation can be rewritten in the following  multiple power representation:
\begin{equation}
\label{MCre}
 D^\text{NS} \cdot C_\text{NS}^\text{Bjp} = \1 + \frac{\beta(a_s)}{a_s}\,\cdot K(a_s)=\1
 +\frac{\beta(a_s)}{a_s}\cdot\sum_{n\geq 1}\left(\frac{\beta(a_s)}{a_s}\right)^{n-1}{\cal
 P}_n(a_s)\,,
\end{equation}
where ${\cal P}_n(a_s)$ are the polynomials in $a_s$ that can be expressed only in terms of the
elements $d_k[\ldots],~c_k[\ldots]$.
In this sense ${\cal P}_n$ do not depend on the $\beta$-function,
all the charge renormalizations being accumulated by  $\left(\beta(a_s)/a_s\right)^{n}$.
Below we present the first two polynomials (factor out $-c^\text{NS}_1=d^\text{NS}_1=3C_\text{F}$)
\begin{subequations}
\label{eq:RHS-CR}
\begin{eqnarray}
{\cal P}_1(a_s)&=&-a_s
d^\text{NS}_1\bigg\{d_2[1]-c_2[1]+a_s\left[d_3[1]-c_3[1]-d^\text{NS}_1(d_2[1]+c_2[1])\right] \\
&&+a_s^2\Big[d_4[1]-c_4[1]-d^\text{NS}_1\left(d_3[1]+c_3[1]+d_2[0]c_2[1]+d_2[1]c_2[0]\right)\Big]\bigg\}\,,\\
{\cal P}_2(a_s)&=&a_s d^\text{NS}_1 \bigg\{d_3[2]-c_3[2]+a_s \Big[
d_4[2]-c_4[2]+ d^\text{NS}_1\left(c_3[2]+d_3[2]\right)\Big]\bigg\}\,,
\end{eqnarray}
\end{subequations}
which were obtained and verified in N$^3$LO in \cite{Kataev:2010du,Kataev:2011zh}
in another normalization.
The construction of the $\beta$-term in the RHS of (\ref{MCre})
also creates constraints for combinations of the $\beta$-expansion elements.
A few chains of these constraints were obtained in \cite{Kataev:2010du}.
Further we shall use the relation
\begin{eqnarray}
  \label{eq:k_n1-d_n1}
 d_2[1] - c_2[1] =d_3[0,1] - c_3[0,1]= \ldots  
= d_n[\underbrace{0,0,\ldots, 1}_{n-1}] - c_n[\underbrace{0,0,\ldots, 1}_{n-1}] =
 \left(\frac{7}2-4\zeta_3 \right),
 \end{eqnarray}
that corresponds to Eq.(30) in \cite{Kataev:2010du}.

If the terms $c_3[1],~d_3[1]$ and $c_4[2],~d_4[2]$ are missed in the $\{\beta\}$-expansion of  $D^\text{NS}$ and $C_\text{NS}^\text{Bjp}$,
as in the variant of the expansion in
\cite{Brodsky:2011ta,Mojaza:2012mf,Wu:2013ei,Brodsky:2013vpa},
the structure of the generalized CR in (\ref{eq:RHS-CR})
is corrupted. That structure certainly contradicts the explicit results of analytical calculations
of  $D^\text{NS}(a_s)$ and  $C_\text{NS}^\text{Bjp}(a_s)$, preformed in the N$^2$LO in
\cite{GorKatL,SurgSam,Chetyrkin:1996ez} and in N$^3$LO in \cite{Baikov:2010je}.

\subsection{ Nonsinglet parts of $D$ and $C^\text{Bjp}$}
\label{sec:rev-exp.1}
Following the approach discussed in  Sec. \ref{sec:expantion structure}
and taking into account a certain definition of the $\beta$-function coefficients in
Eq.~(\ref{eq:beta})
we can obtain the $\beta$-expansion for $D$-- and $C$--functions.
For the Adler function $D^\text{NS}$ it reads
\begin{subequations}
\label{eq:d1-4}
 \begin{eqnarray}
d_1^\text{NS}&=&3{\rm C_F};~d_1=~1; \label{D-11}\\
d_2[1]&=&\frac{11}2-4\zeta_3;~~~~~
d_2[0]=\frac{\rm C_A}3-\frac{\rm C_F}2=\frac{1}{3}; \label{D-21} \\
d_3[2]&=&\frac{302}9-\frac{76}3\zeta_3\approx 3.10345;~d_3[0,1]=\frac{101}{12}-8\zeta_3\approx
-1.19979;\label{D-32}\\
d_3[1]&=&
    {\rm C_A}\left(\frac{3}4 + \frac{80}3\zeta_3 -\frac{40}3\zeta_5\right) -
    {\rm C_F}\left(18 + 52\zeta_3 - 80\zeta_5\right)\approx 55.7005 \label{D-31}; \\
d_3[0]&=& \left(\frac{523}{36}-
        72 \zeta_3\right){\rm C_A^2}
    +\frac{71}3 {\rm C_A C_F} - \frac{23}{2} {\rm C_F^2}\approx -573.9607~,  \label{D-30}
\end{eqnarray}
\end{subequations}
which differs from the ones presented in Ref. \cite{Kataev:2010du}
(see its ``natural form'' in the Appendix \ref{App-B}),
by the normalization factor only.
It  looks more convenient for a certain BLM task
(the presentation corresponds to one in \cite{MS04}) due to setting of the first PT
coefficient, $d_1~(c_1)$, equal to 1.
Let us emphasize that gluinos are used here as a pure technical device
to reconstruct the $\beta$-function expansion of the perturbative coefficients.

In this connection, we mention the relation $d_3[0,1]=d_n[0,\ldots,1]=d_2[1]$
proposed in \cite{Mojaza:2012mf} and based on  a ``special degeneracy of the coefficients''
suggested there (see Eq.~(6) in \cite{Mojaza:2012mf})
in an analogy with the perturbative series rearrangement, $d_i \to d'_i$ under the change of
the coupling renormalization scale,
$a(\mu^2) \to a'(\mu'^2)$
(see below the discussions around  Eq.~(\ref{eq:rearrange})).
This rearrangement has an \textit{outside} reason with respect to $d_i$ and ``does not know''
about the \textit{intrinsic} structure of the initial coefficient $d_i$  under consideration.
This relation is artificial and by this reason it is not supported by the direct calculations.
The explicit result of this rearrangement is presented in Eq.~(\ref{eq:rearrange}),
it is the initial step for any BLM optimization procedure that
will be discussed in Sec.\ref{sec:optBLM} in detail.

Let us compare now Eqs.~(\ref{eq:d1-4}) with the results presented in \cite{Brodsky:2013vpa}
and based on the interpretation of the term
$\left(\underline{48\gamma_3} +3\beta_1 \Pi_1+\underline{24\beta_0 \Pi_2}\right)a_s^3$
in the presentation of (\ref{Adler:master:exp}).
The first and the third terms of the sum form the term proportional to $\beta_0^2$
\ba
\underline{48\gamma_3}+ \underline{24\beta_0 \Pi_2} \longrightarrow \beta_0^2
\left(\frac{302}9-\frac{76}3\zeta_3=d_3[2] \right)
\ea
 that can be unambiguously obtained by
 extracting the $n_f^2$ terms in  $\gamma_3$ and the $\beta_0$-term in $\Pi_2$, see
 the corresponding explicit expressions in \cite{Baikov:2012zm}.
 The second term there, $\beta_1 \Pi_1$, certainly contributes to the value of the element
 $d_3[0,1]$.
 There are other terms, proportional to $\beta_1$, in both the $\gamma_3$ and $\beta_0 \Pi_2$
terms that also contribute to $d_3[0,1]$.
 However,  these required terms cannot be separated unambiguously from those terms that are
 proportional to $\beta_0$.
 The final explicit expressions given in \cite{Baikov:2012zm} are not sufficient for this
 separation,
 as it was already discussed in Subsec.\ref{subsec:Explicit D}.

Let us consider the $\beta$-expansion of the Bjorken coefficient function
$C_{\rm NS}^\text{Bjp}$ of the DIS sum rules.
Based on CR (\ref{eq:CI-PT0})
for  $n=2$ and $n=3$    and the already fixed $d_2[0]$ and $d_3[0]$-terms
we get expression (\ref{eq:c3-d3})  for the $c_2[0]$ and
$c_3[0]$ elements of $C_{\rm NS}^\text{Bjp}$,
namely, $c_3[0] =  d_3[0]-2d^\text{NS}_1 d_2[0]+(d^\text{NS}_1)^2$,
see the explicit expression in (\ref{C-30}).
The knowledge of $c_3[0]$ allowed us to fix all  other elements
$c_3[\ldots]$   of the PT coefficient $c_3$
without involving additional  degrees of freedom \cite{Kataev:2010du}.
It is instructive to consider this in detail.
Indeed, the terms $c_3[0]$ as well as the coefficient $c_3[2]$ of the $\beta_0^2$
(maximum power of $n_f$) can be found independently.
Therefore, the Casimir structure of the rest of $c_3$,
$c_3-c_3[0]-\beta_0^2c_3[2]$,
contains \textbf{5} basis elements (we factor out $c^\text{NS}_1=-3C_\text{F}$):
$$c_3-c_3[0]-\beta_0^2c_3[2]:
\left\{
  \begin{array}{l}
 ~ C^2_\text{F},~C^2_\text{A},~C_\text{A}C_\text{F}, ~T_\text{R}n_f C_\text{F},
~T_\text{R}n_f C_\text{A}\\
~ \beta_1,~\beta_0\\
\end{array}
  \right.
 $$
This Casimir expansion of the rest should be equated to the $\beta$-expansion of the one
(see decomposition (\ref{eq:d_beta})),
 $c_3[0,1] \cdot \beta_1+ \left(x \cdot C_\text{F} + y\cdot C_\text{A}\right)\cdot \beta_0$
 that contains 3 unknown coefficients $c_3[0,1],x,y$.

The $C^2_\text{F}$-terms in the explicit result for $c_3$ \cite{Larin:1991tj}
(see App.\ref{App:A}, Eq.(\ref{eq:C1-3})) and in the expression for $c_3[0]$ in
Eq.(\ref{C-30}) coincide to one another;
therefore,  the term $\frac{1}2C^2_\text{F}$ is canceled  in the rest.
This confirms the fact that its $\beta$-expansion does not contain $C^2_\text{F}$.
So we have \textbf{4} constraints (not \textbf{5}) for the \textbf{3} coefficients $c_3[0,1],x,y$.
This overdetermined system  is a system of simultaneous equations;
the fact provides us with an \textit{independent confirmation} of this $\beta$-expansion.
The explicit form of the elements were first obtained in \cite{Kataev:2010du};
below we present them at the same normalization
as Eq.~(\ref{eq:d1-4}) (cf. (\ref{D-30}) with (\ref{C-30})):
\begin{subequations}
 \label{eq:c1-4}
\begin{eqnarray}
c^\text{NS}_1&=&-3~{\rm  C_F};~c_1=1; \label{c-11}\\
c_2[1]&=& 2;
c_2[0]= \left(\frac{1}{3}{\rm C_A}-\frac{7}{2}{\rm C_F}\right)=-\frac{11}3=-3.6(6);
 \label{c-21}
\end{eqnarray}
\begin{eqnarray}
c_3[2]&=& \frac{115}{18}=6.38(8);~c_3[0,1]=\bigg(\frac{59}{12}-4\zeta_3\bigg)\approx 0.10844;
\label{C-32} \\
c_3[1]&=& -\bigg(\frac{166}{9}- \frac{16}3\zeta_3\bigg){\rm C_F} -
\bigg(\frac{215}{36}- 32 \zeta_3+\frac{40}{3}\zeta_5\bigg){\rm C_A}\approx 39.9591;\label{C-31} \\
c_3[0] &=&
\bigg(\frac{523}{36} - 72\zeta_3\bigg){\rm C_A^2}+\frac{65}{3}{\rm C_F C_A}+ \frac{\rm C_F^2}{2}
\approx -560.627. \label{C-30}
\end{eqnarray}
\end{subequations}
The same results  can be obtained if one fixes first the element
$c_3[0,1]= d_3[0,1] -d_2[1]+c_2[1]$ from
relation (\ref{eq:k_n1-d_n1}), the latter originates from another source --
the symmetry breaking term proportional to $\beta(a_s)$ in the generalized CR.
Therefore,
the results (\ref{eq:c1-4}) are in mutual agreement with both the terms
in the RHS of CR and can be obtained independently from each of them.

These  elements of decomposition in (\ref{eq:c1-4}) allows one to make a \textit{new prediction}
for the light gluino contribution to $C_\text{NS}^{\rm Bjp}$.
Indeed,
for the considered here RGI quantities the effects of charge renormalization appear in two ways:
the elements $c[...]$ -- the coefficients of the $\beta$-function products
(named $B^j$ in \ref{subsec:Formulation}) -- are formed following gauge interaction;
the effect of various degrees of freedom, say, gluino,
which reveals itself only in intrinsic loops,
changes the content of the $\beta$-coefficients $\beta_i$ with the corresponding mark,
say, $\bm{n_{\tilde{g}}}$.
Therefore, to obtain $C^{\rm Bjp} \to C^{\rm Bjp}(a_s, n_f, n_{\tilde g})$ with the
light MSSM gluino one should replace the $\beta$-coefficients
$\beta_i \to \beta_i\left(n_f, n_{\tilde{g}}\right)$ and compose them
with the elements from Eq.(\ref{eq:c1-4}),
\begin{subequations}
 \label{eq:c1-3beta}
 \begin{eqnarray}
&&C_\text{NS}^{\rm Bjp}(a_s, n_f, n_{\tilde g}) =1+  a_s(-3C_F)  \\
&&+a_s^2(-3C_F)\cdot
          \left\{\frac{1}{3}{\rm C_A}-\frac{7}{2}{\rm C_F}+
          2\beta_0\left(n_f, n_{\tilde{g}}\right) \right\} \label{c3:gl} \\
&&+a_s^3(-3C_F)\cdot
          \left\{\frac{115}{18}\beta_0^2\left(n_f, n_{\tilde{g}}\right)+
          \left(\frac{59}{12}-4\zeta_3\right)\beta_1\left(n_f, n_{\tilde{g}}\right)\right. \nonumber
          \\
&&        \left.  -\left[\left(
\frac{215}{36}- 32 \zeta_3+\frac{40}{3}\zeta_5\right){\rm C_A}
    +\bigg(\frac{166}{9}- \frac{16}3\zeta_3\bigg){\rm C_F}\right]\beta_0\left(n_f,
    n_{\tilde{g}}\right)\right.
    \nonumber \\
&&     \left. +
 \bigg( \frac{523}{36} - 72\zeta_3\bigg){\rm C_A^2}+\frac{65}{3}{\rm C_F C_A}+ \frac{\rm C_F^2}{2}
 \right\}.
 \end{eqnarray}
  \end{subequations}
This logic can be reverted: the values of $c_3[0],~c_3[0,1]$ and then the CR can be checked
from the direct calculation of $C_\text{NS}^{\rm Bjp}(a_s, n_f, n_{\tilde g})$
with the MSSM massless gluino.

\subsection{Singlet parts and the $R$-ratio}
 \label{sec:$R$-ratio}
Here we present the singlet part of the Adler function, $d_4$, that can be obtained based on the
result for $c_4$
of $C_\text{S}^{\rm Bjp}$
and CR \cite{Baikov:2012zn},
\ba
d_4^S&=& \beta_0(n_f)\cdot d_{4}^{S}[1]+d_{4}^S[0]\ ,
\ea
\ba
d^S_{4}[0]&=&
\left(-\frac{13 }{64}\zeta_3-\frac{5 }{32}\zeta_5+\frac{205}{1536}\right)
C_A+\left(-\frac{1}{4}\zeta_3+
\frac{5}{8} \zeta_5-\frac{13}{64}\right) C_F \ , \\
d^S_{4}[1]&=&-\frac{13}{32} \zeta_3-\frac{1}{8}\zeta_3^2+\frac{5}{16}\zeta_5+\frac{149}{576}\ ,
\ea
\ba
c_4^S&=& c_{4}^S[0]+\beta_0(n_f)\cdot c_{4}^{S}[1]\ , \\
c_{4}^S[0]&=&
\left(\frac{13 }{64}\zeta_3+\frac{5 }{32}\zeta_5-\frac{205}{1536}\right) C_A+
\left(\frac{1}{16}\zeta_3-\frac{5 }{8}\zeta_5+\frac{37}{128}\right) C_F\ , \\
 c_{4}^S[1] &=&-\frac{119}{1152}+\frac{67}{288}
 \zeta(3)+\frac{1}{8}\zeta(3)^2-\frac{35}{144}\zeta(5)\,
\ea
 The integral transform $D \to R_{e^+e^-}$,
\begin{eqnarray}
 \label{eq:D->R}
   R_{e^+e^-}(s)&\equiv &   R(s,\mu^{2}=s)=
  \frac{1}{2\pi \textit{i}}
  \int_{-s-i\varepsilon}^{-s+i\varepsilon}\!
  \frac{D^{\rm EM}(\sigma/\mu^2;a_s(\mu^2))}
  {\sigma}\,
  d\sigma~
  \Bigg|_{\mu^2=s} =  \\ \nonumber
  &=& \left(  \sum_i q_i^2 \right) d_R \left (1+ \sum_{m\geq 1} r_m^{\rm NS} \, a_s^m(s)\right )+
  \left(  \sum_i q_i \right)^2 \frac{d^{abc}d^{abc}}{d_R} \sum_{n\geq 3} r_{n}^{\rm S} a_s^{n}(s)\,,
\label{eq:FOPT}
\end{eqnarray}
 can be realized as a linear relation by means of the matrix $T$,
$ r_j= T^{j i}d_{i}$, or for the vector representation $R=T D= \sum a_s^j T^{j i}d_{i}$.
The triangular matrix $T$ of the relation can be obtained at any fixed order of perturbative theory
\cite{BMS2010jhep}.
The elements of this matrix below the units on the diagonal contain so-called kinematic
``$\pi^2$-terms''
multiplied by the $\beta$-function coefficients\footnote{
These terms can be obtained for any order of perturbative theory
(constrained  mainly by the value of RAM) with ``Mathematica'' routine constructed
by V. L. Khandramai and S. V. Mikhailov.}, 
see an example of $T^{j i}$ in Appendix \ref{App-D}.
Taking into account that the  $\beta$-structure of the normalized coefficients
$r_i=r^\text{NS}_i/r^\text{NS}_1$  is like that for the coefficients $d_i$, Eqs.(\ref{eq:d_beta}),
one can rewrite the results from the matrix in Table~\ref{Tab:r_n.d_k},
 \begin{subequations}
\label{eq:D-R}
\begin{eqnarray}
\label{eq:r_2}
&&r^{}_0=d^{}_0;~r^\text{NS}_1=d^\text{NS}_1;~r^{}_1=1\,;~r^{}_2=d^{}_2;\\
&&r_3[2]=d_3[2]\BluTn{-\frac{\pi^2}3 };~
r^{\rm S}_3[2]=d^{\rm S}_3[2];
\label{eq:r_3} \\
&&r_4[3]
= d_4[3]\BluTn{- \pi^2 d_2[1]};~r_4[2]=d_4[2]\BluTn{- \pi^2 d_2[0]};
~r_4[1,1]=d_4[1,1]\BluTn{- \frac{5}6\,\pi^2};      \label{eq:r_4} \\
&&r_5[4]= d_5[4]\BluTn{+ \frac{\pi^4}5 -  2 \pi^2d_3[2]};
r_5[0,2]=d_5[0,2]\BluTn{- \frac{\pi^2}2}; r_5[2,1]=d_5[2,1]\BluTn{-\pi^2\left( \frac{7}3d_2[1]+
d_3[0,1]\right)}; \nonumber \\
&&r_5[1,0,1]=d_5[1,0,1]\BluTn{- \pi^2};r_5[1,1]=d_5[1,1]\BluTn{-
\frac{7}3\pi^2d_2[0]};r_5[3]=d_5[3]-\BluTn{2 \pi^2 d_3[1]}; \nonumber \\
&&r_5[2]=d_5[2]-\BluTn{2 \pi^2 d_3[0]}, \label{eq:r_5}
\end{eqnarray}
 \end{subequations}
while the other elements in $r_i$  coincide with ones in $d_i$ ($i\leq 5$).
A similar matrix $T_{nl}^{S}$  which relates the coefficients
$r_n^{\rm S}$ and $d_l^{\rm S}$, can be constructed as well. However, in this
work we will not consider the $\pi^2$-dependent   effects of
analytical continuation, which  in the singlet case appear first
at the ${\cal O}(a_s^5)$-level.
Further, we shall use Eqs.(\ref{eq:D-R}) to construct PT optimized series for $R$.

\section{BLM and PMC procedures and the results}
 \label{sec:optBLM}
\subsection{ General basis}
 The re-expansion of  the running coupling $\bar{a}(t)= a(\Delta, a')$
and its powers
in terms of $t-t'=\Delta=\ln\left(\mu^2/\mu'^2\right)$ and new coupling $a'$ reads,
\ba
\label{eq:RGrearrange}
\bar{a}(t) = a(\Delta, a')&=& a' - \beta(a')\frac{\Delta}{1!} +
\beta(a')\partial_{a'}\beta(a')\frac{\Delta^2}{2!}
+\beta(a') \partial_{a'} \left( \beta(a')\partial_{a'}\beta(a') \right)\frac{\Delta^3}{3!} + \ldots
\nonumber \\
&=& \exp\left(-\Delta \beta(\bar{a})\partial_{\bar{a}}\right)\bar{a}\mid_{\bar{a}=a'},
\ea
which is the way to write the corresponding RG solution for $a(t)$
through the operator  
$\exp\left(-\Delta \beta(a)\partial_{a}\right)[\ldots]\mid_{a=a'}$
 (see \cite{MS04} and refs therein).
The shift of the logarithmic scale
$\Delta$ in its turn can be expanded
in perturbative series in powers of $a'\beta_0$
\ba \la{Delta}
~t' &\equiv & t-\Delta, \nonumber \\
&& \Delta\equiv\Delta(a')=\bm{\RedTn{\Delta_{0}}} + a'\beta_0 \bm{\RedTn{\Delta_{1}}}
+ (a'\beta_0)^2 \bm{\RedTn{\Delta_{2}}} + \ldots,
\ea
 where the argument of the new coupling $a'$ depends on $t'=t-\Delta$.
 It is sufficient to take this renormalization scale for the $a'$ argument, which corresponds to the
 solution
on the previous step, rather than  to solve the exact equation  $a\left(t-\Delta(a')\right)=a'$.
Re-expansion  $a$ in terms of $a'$ leads to rearrangement of the series of perturbative expansion for
the
RGI quantity $D^a~(C^\text{Bjp})$,
$a^i d_i \to a'^i d'_i$, where the r.h.s. are expressed in a rather long but evident formulae.
n the square brackets below, we write them explicitly:
\begin{subequations}
\label{eq:rearrange}
  \ba
a^1\cdot d_1 \to&a'^1\cdot [d'_1=& 1]; \nonumber  \\
a^2\cdot d_2 \to&a'^2\cdot \big[d'_2=&\beta_0\,d_2[1]
  + d_2[0] -\beta_0 \bm{\RedTn{\Delta_{0}}}\big]; \la{1-stage-2}
\ea
 \ba
a^3\cdot d_3 \to&a'^3\cdot \Big[d'_3=&\beta_0^2 \left(d_3[2]- 2d_2[1]\bm{\RedTn{\Delta_{0}}}+
\bm{\RedTn{\Delta_{0}^2}}\right)+
\beta_1\left(d_3[0,1]-\bm{\RedTn{\Delta_{0}}}\right)+ \\
&&\beta_0 \left( d_3[1]-2d_2[0]\bm{\RedTn{\Delta_{0}}}\right)+d_3[0]-
\beta_0^2 \bm{\RedTn{\Delta_{1}}} \Big];
 \la{1-stage-3}\\
a^4\cdot d_4 \to&a'^4\cdot\Big[d'_4=&\beta_0^3\left(d_4[3] -3d_3[2]\bm{\RedTn{\Delta_{0}}}+
3d_2[1]\bm{\RedTn{\Delta_{0}^2}}-\bm{\RedTn{\Delta_{0}^3}}-
2\left(\bm{\RedTn{\Delta_{0}}}-d_2[1]\right)\bm{\RedTn{\Delta_{1}}}\right)+
\nonumber \\
&&\beta_1\beta_0\left(d_4[1,1]-\left(3d_3[0,1]+2d_2[1] \right)\bm{\RedTn{\Delta_{0}}}
+\frac{5}2\bm{\RedTn{\Delta_{0}^2}}-\bm{\RedTn{\Delta_{1}}}\right)
+ \nonumber \\
&&\beta_2\left(d_4[0,0,1]- \bm{\RedTn{\Delta_{0}}}\right)+ \label{1-stage-3}\\
&&\beta_0^2 \left(d_4[2] -3d_3[1]\bm{\RedTn{\Delta_{0}}}+
3d_2[0]\bm{\RedTn{\Delta_{0}^2}}-
2d_2[0]\bm{\RedTn{\Delta_{1}}}\right)+\nonumber \\
&&\beta_1\left(d_4[0,1]-2d_2[0]\bm{\RedTn{\Delta_{0}}}\right)+ \\
&&\beta_0\left(d_4[1]-3d_3[0]\bm{\RedTn{\Delta_{0}}}\right)+ d_4[0]-\beta_0^3\bm{\RedTn{\Delta_{2}}}
\Big];      \\
\ldots&&\ldots \nonumber \\
a^{n}~\cdot d_{n} \to &a'^n~\cdot \Big[d'_n=&\beta_0^{n-1}d_n[n-1]+  \ldots \Big]\,.
\label{1-stage-n}
\ea
 \end{subequations}
The standard BLM fixes the scale $\bm{\RedTn{\Delta_{0}}}$ by the requirement
$\bm{\RedTn{\Delta_{0}}}=d_2[1]$,
 accumulating 1-loop renormalization of charge just in this new scale \cite{Brodsky:1982gc},
at the same time the coefficient $d_2 \to d_2[0]$ -- its ``conformal part".
Numerically,
\be
 \bm{\RedTn{\Delta_{0}}}=d_2[1] =\frac{11}2-4\zeta_3 = 0.69177\ldots \approx \ln(2)=0.69314\ldots,
  \ee
therefore, $Q^2_{BLM}=\exp(-\bm{\RedTn{\Delta_{0}}})Q^2\approx Q^2/2$.

High order generalization of BLM can be realized in different ways
requiring consequently certain equations for the partial shifts $\{\bm{\RedTn{\Delta_{i}}} \} $.
The system of Eqs.~(\ref{1-stage-2}-\ref{1-stage-n}) for $d'_i$ is the basis to construct
different BLM generalizations.
It is instructive to consider these coefficients $\{d'_i\}$ after the first BLM step;
taking $\bm{\RedTn{\Delta_{0}}}=d_2[1]$ one obtains
 \begin{subequations}
\label{eq:1BLM step}
\ba
d'_2&=&d_2[0]; \label{2-stage-2}\\
d'_3&=&\beta_0^2\underline{(d_3[2]- d_2[1]^2)}+
\beta_1\underline{\underline{\left(d_3[0,1]-d_2[1]\right)}}+
\nonumber \\
&&\beta_0 \left( d_3[1]-2d_2[0]d_2[1]\right)+d_3[0]-\beta_0^2 \bm{\RedTn{\Delta_{1}}};
\label{2-stage-3}
\ea
\ba
d'_4&=&\beta_0^3\underline{\left(d_4[3] -3d_3[2]d_2[1]+2d_2[1]^3\right)}+
 \beta_2\underline{\underline{\left(d_4[0,0,1]- d_2[1]\right)}} \nonumber \\
&&\beta_1\beta_0\left(d_4[1,1]-3d_3[0,1]d_2[1]
+d_2[1]^2/2-\bm{\RedTn{\Delta_{1}}}\right)
+ \label{2-stage-3c}\\
&&\beta_0^2 \left(d_4[2] -3d_3[1]d_2[1]+
3d_2[0]d_2[1]^2-
2d_2[0]\bm{\RedTn{\Delta_{1}}}\right)+ \label{2-stage-4c}\\
&&\beta_1\left(d_4[0,1]-2d_2[0]d_2[1]\right)+ \\
&&\beta_0\left(d_4[1]-3d_3[0]d_2[1]\right)+ d_4[0]-\beta_0^3\bm{\RedTn{\Delta_{2}}};
\label{2-stage-4}    \\
\ldots&&\ldots \nonumber \\
d'_n&=&\beta_0^{n-1}d_n[n-1]+  \ldots  \label{2-stage-n}
\ea
 \end{subequations}
The detailed analysis of the $d'_i$ structure was made in \cite{MS04} in Sec.5.
Here we mention a common  property of this transform --
to obtain the rearrangement of the coefficient at an order $n+1$, $d_{n+1}\to d'_{n+1}$,
one should know its $\beta$-structure up to the previous order $n$.
For the partial case of relation $d_{n+1}[n]=\left(d_2[1]\right)^{n}$ the
$\beta_0^{n}$-terms are canceled (underlined terms) in all the orders even due to the first BLM
step.
Correspondingly, the special conditions $d_i[0,\ldots,1]=d_2[1]$ will remove the next terms with the
leading coefficient $\beta_{i-2}$ in every order,
see double underlined terms in Eqs.~(\ref{2-stage-3},\ref{2-stage-3c}).
The latter conditions were proposed  in \cite{Mojaza:2012mf}
(see the discussion in Sec.\ref{sec:rev-exp.1} after Eq.~(\ref{eq:d1-4})),
though both of the above hypotheses  are far from the results of the direct calculations at ${\cal
O}(a_s^3)$
in (\ref{eq:d1-4}), really
\ba
d_3[2]-d_2[1]^2 \approx 3.1035-0.4785;
~~~d_3[0,1]-d_2[1] \approx 1.1998-0.6918 \,.
\ea
Even more, in QCD the elements $d_{n+1}[n]$ grow as $n!$ due to renormalon contributions \cite{BK93}
and the role of these terms becomes more and more important.
To construct the next steps of the PT-optimization with
$\bm{\RedTn{\Delta_{1}}}, \bm{\RedTn{\Delta_{2}}}, \ldots$, one should get more detailed knowledge
or
provide a hypothesis about the different contributions to $d'_n$.
   \subsection{seBLM and PMC procedures}
 \label{sec:seBLMPMC}
One of the hypotheses mentioned above is based on the empirical relation between the QCD
$\beta$--function coefficients $\beta_i$, ~$\beta_i \sim \beta_0^{i+1}$.
This can be easily verified for  perturbative quenced QCD
($n_f=0$) numerically and this works in the range of $n_f=0\div 5$ of quark flavors
for the all known $\beta$-coefficients;
compare the expressions in Eqs.~(\ref{eq:beta0-3},\ref{eq:beta3}),
\be \label{eq:c_i}
  \beta_i \sim \beta_0^{i+1}, c_i= \beta_i/\beta_0^{i+1}={\cal O}(1).
  \ee
This relation  allows one to set
a hierarchy of contributions
 in order of the ``large value of $\beta_0$'' ($\beta_0=11 (9)$  at $n_f=0 (3)$) \cite{MS04}.
Of course, relation (\ref{eq:c_i}) should be broken at some large enough order of expansion $i_0$
 in virtue of expected Lipatov like asymptotics for the $\beta$-function $\beta_i \sim
 (i!)\beta_0^{i+1}$.
 Therefore, this hierarchy has a restricted field of application that describes the term
 ``practical approach'' in the title of \cite{MS04}.

For this hierarchy the most important terms are of an order of $(\beta_0a_s)^n/\beta_0$
  in order $n$ -- underlined below in Eq.~(\ref{eq:R_se}).
  For illustration we shall use the $R^\text{NS}(s)$-ratio taking into account the result
  (\ref{eq:1BLM step}) for $D$ and relations  in Eq.~(\ref{eq:D-R})
\begin{subequations}
 \label{eq:R_se}
  \begin{eqnarray}
  \label{eq:1-stage-R_2}
r'_2 &=&\! d_2[0]\, ,\\
  r'_3&=&\!
 \beta_0^2\underline{(d_3[2]- d_2[1]^2 \BluTn{-\pi^2/3} )}+
\beta_1\underline{\left(d_3[0,1]-d_2[1]\right)} \\
              && + \underline{ \underline{\beta_0 \left( d_3[1]-2d_2[0]d_2[1]\right)}}
  -\beta_0^2 \bm{\RedTn{\Delta_{1}}}+d_3[0]; \label{1-stage-R3c}
 \end{eqnarray}
  \ba
  r'_4 &=&\beta_0^3\underline{\left(d_4[3] -3d_3[2]d_2[1]+2d_2[1]^3\BluTn{- \pi^2 d_2[1]}\right)}+
 \beta_2\underline{\left(d_4[0,0,1]- d_2[1]\right)} \nonumber \\
&&\beta_1\beta_0
\left(\underline{d_4[1,1]-3d_3[0,1]d_2[1]+d_2[1]^2/2\BluTn{-
5/6\,\pi^2}-\bm{\RedTn{\Delta_{1}}}}\right)
+ \label{1-stage-R34d}\\
&&\beta_0^2 \left(\underline{\underline{d_4[2] -3d_3[1]d_2[1]+
3d_2[0]d_2[1]^2 \BluTn{- \pi^2 d_2[0]}-
2d_2[0]\bm{\RedTn{\Delta_{1}}}}}\right)+ \label{1-stage-R4e}\\
&&\beta_1\left(\underline{\underline{d_4[0,1]-2d_2[0]d_2[1]}}\right)+ \label{1-stage-R4f}\\
&&\beta_0\left(\underline{\underline{\underline{d_4[1]-3d_3[0]d_2[1]}}}\right)
-\beta_0^3\bm{\RedTn{\Delta_{2}}}+ d_4[0];  \label{1-stage-R4g}    \\
 r'_{n}
   &=&\! \! \! \! \! \!~~\underline{\beta_0^{n-1}\! d_{n}[n\!-\!1]+ \ldots}+  \ldots
\label{eq:d_nse}
\end{eqnarray}
 \end{subequations}
The less important terms are suppressed by $\beta_0^{-1}$ in this order;
$(\beta_0a_s)^n/\beta_0^2$, they are double-underlined in
(\ref{1-stage-R3c},\ref{1-stage-R4e},\ref{1-stage-R4f}), and so on.
Following the hierarchy one fixes the values of
$\bm{\RedTn{\Delta_{1}}}, \bm{\RedTn{\Delta_{2}}}, \ldots$ consequently
nullifying at first the most important (1-underlined) $\beta$-terms in every order.
After that the procedure repeated with the less important terms (double underlined)
 in all orders, etc.
 This procedure was called sequential BLM (seBLM) and its
 result was presented in detail in Sec.~6 of \cite{MS04} (see Eqs.~(6.7,6.8) there).
The discussed  hierarchy can also be used for generalization of the NNA approximation,
 see Appendix C in \cite{BMS2010jhep}.

The above invented hierarchy ignores a possible difference of values of the elements $d_n[\ldots]$
tacitly suggesting that they are of the same order of magnitude.
Of course, one can abandon the suggestion of the hierarchy,
 and to remove \textit{all the $\beta$-terms in one mold } consequently order by order.
  This approach leads to other values of $\bm{\RedTn{\Delta_{i}=\bar{\Delta}_{i}}}$:
\begin{subequations}
\label{eq:Delta-012}
 \ba
  &&\bm{\RedTn{\bar{\Delta}_{0}}} = d_2[1];\label{Delta(0)}\\
&&\bm{\RedTn{\bar{\Delta}_{1}}} = \frac{1}{\beta_0^2}\left[\beta_0^2 \left(d_3[2]-
d^2_2[1]-\BluTn{\pi^2/3 }\right)+\beta_1\left(d_3[0,1] - d_2[1]\right)+
   \beta_0 \underline{\underline{\left( d_3[1]-2d_2[1]d_2[0]\right)}}\right] \label{Delta(1)} \\
&&\bm{\RedTn{\bar{\Delta}_{2}}}= \frac{1}{\beta_0^3}\left[
\beta_0^3\left(d_4[3]-3d_2[1]d_3[2]+2(d_2[1])^3 \BluTn{-\pi^2 d_2[1]}\right)
+\beta_2 (d_4[0,0,1]-d_2[1])+
\right. \nonumber \\
&&\left.\phantom{\Delta_{0} = } \beta_0 \beta_1
\left(d_4[1,1]-3d_3[0,1]d_2[1]+\frac3{2}(d_2[1])^2 - d_3[2] -\BluTn{\pi^2/2}\right)+
\beta_1^2/\beta_0 \left(d_2[1]- d_3[0,1]\right) +\right. \nonumber \\
&&\left. \phantom{\Delta_{0} = } \beta_1\left(\underline{\underline{\ldots}}\right)+\ldots
\right]\,,
\label{Delta(2)}
\ea
 \end{subequations}
which differ by the underlined ``suppressed in the $1/\beta_0$'' terms from the previous ones in
\cite{MS04}.
The complete form for $\Delta_2$ looks cumbersome and it is outlined in Appendix~\ref{App-E}.
 The procedure like this was called PMC later on \cite{Brodsky:2011ig},
  though for both the cases, seBLM and
  corrected PMC, the final PT series
   has the same ``conformal terms'' $d_n[0]$  as the coefficients of new expansion.
The new normalization scale $s'$ follows from Eq.~(\ref{Delta}),
taking into account certain expressions for $\Delta_i$ in (\ref{eq:Delta-012}),
\begin{subequations}
 \label{eq:PMC2}
 \ba
  \label{eq:PMC2a}
 R_{e^+e^-}(s) &=&\left(\sum_i q_i^2 \right)\cdot d_{A} R^{\rm NS} +
 \left(\sum_i q_i \right)^2 \,\cdot d_{A} R^\text{S} \nonumber \\
 R^\text{NS}(s)&=& 1 + 3C_F \left\{a(s')
  + d_2[0] \cdot a^2(s') +d_3[0]\cdot a^3(s') + d_4[0]\cdot a^4(s')+ \ldots\right\} \label{eq:PMC2a}
  \\
  \ln(s/s')&=& \bar{\Delta}_{0} + a'\beta_0 \bar{\Delta}_{1}
+ (a'\beta_0)^2 \bar{\Delta}_{2}+ \ldots \,. \label{eq:PMC2b}
\ea
 \end{subequations}
 The formulae Eqs.(\ref{eq:Delta-012}) and  Eqs.(\ref{eq:PMC2}) are the main results
 of these subsections.

\subsection{Numerical estimates, discussion of PMC/seBLM results}
Here we apply the results of the procedure accumulated in
Eqs.~(\ref{eq:R_se},\ref{eq:Delta-012},\ref{eq:PMC2})
 for the numerical estimates of the expansion coefficients for a few processes
 starting with the nonsinglet part $R^\text{NS}$ of the $R_{e^+e^-}(s)$-ratio.
The corresponding singlet part $R^\text{S}$ can be optimized independently; moreover
it is not very important numerically.
For the sake of illustration, we put the value $n_f=3$ for all estimates below.
At the very beginning we have the following numerical structure of $r_i$,
\begin{subequations}
 \ba
&& r_2 =  \beta_0\cdot 0.69~+~ \frac1{3}\approx 6.56;  \la{R-r2}
\ea
 \ba
&&\Ds r_3 =  -\beta_0^2 \cdot0.186~~ - ~~\beta_1 \cdot 1.2~~~ + \beta_0 \cdot 55.70~-573.96\,
\approx-164.5  \la{R-r3numbers}\\
&&\Ds ~~~~~~~~~~~-15.1~~~~-~~~76.8~~~~+~~501.3~~~~~-
573.96  ~~~~~~~~~~~ \nonumber \\
&&\Ds r_4 \approx -6840.29 \label{R-r4numbers}
\ea
 \end{subequations}
At the first BLM setting $\bar{\Delta}_{0}\approx 0.69 $, $a_s(s) \to a'_s=a_s(s e^{-0.692}\approx
s/2)$ we obtain
for the coefficients $r'_2,r'_3$ --- Eqs.~(\ref{R1-r2},\ref{R1-r3numbers}) --- the explicit result
of the BLM procedure.
The value of the second coefficient $r'_2$ diminishes by an order of magnitude,
while $r'_3$ becomes moderately larger,
compare (\ref{R-r3numbers}) with (\ref{R1-r3numbers},\ref{R2-r3numbers}),
\begin{subequations}
 \label{R1}
\ba
&&r'_2  =  \frac1{3};  \la{R1-r2}\\
&&\Ds r'_3 =  -\beta_0^2 \cdot0.665~~ - ~~\beta_1 \cdot 1.892~~~ + \beta_0 \cdot 55.24~-573.96\,
\approx-\underline{251.7} ;\la{R1-r3numbers}\\
&&\Ds ~~~~~~~~~~~-53.86~~~~-~~~~~121.0~~~~+~~~497.1~~-573.96   ~~~~~~~~~~~\nonumber\\
&&\Ds r'_4 \approx -8559.89\,. \label{R1-r4numbers}
\ea
 \end{subequations}
At the second step (PMC), we obtain $\bar{\Delta}_1 \approx 3.98$ following  Eq.(\ref{Delta(1)}) and
Eq.(\ref{Delta}),
\begin{subequations}
 \ba
 &&  \Ds r''_3 \approx -\underline{573.96}, ~~\bar{\Delta}_1 \approx 3.98\,, \la{R2-r3numbers}\\
 &&\Ds r''_4 \approx -11066.1\,, \label{R2-r4numbers} \\
 && a'_{s} \to a^{''}_s=a_s\left(s\cdot e^{-0.692-3.98 \beta_0 a'_s(s) }\right)\,,
\ea
 \end{subequations}
while $r''_3=d_3[0]$  following  the main aim of PMC, see Eq.~(\ref{eq:PMC2a}).
Due to the strong suppression  of the normalization scale by a factor of $\exp\left[-0.692-3.98
\beta_0 a'_s(s)\right]$
the applicability of PT is shifted to the region of very large $s$;
simultaneously the coefficient $r''_3$ increases 3 times (cf. (\ref{R-r3numbers})).
So this procedure makes the convergence of PT worse.

Within the same framework we obtain for the coefficient of the Bjorken function
$C_\text{NS}^\text{Bjp}$
\begin{subequations}
\label{eq:R-C}
 \ba
&& c_2 =  \beta_0\cdot 2~-~ \frac{11}{3}= 14.3(3);  \la{R-r2}\\
&&\Ds c_3 =  \beta_0^2 \cdot6.39~~ + ~~\beta_1 \cdot 0.1084~~~ + \beta_0 \cdot39.95~-560.63\, \approx
323.44 ;
\la{R-c3numer}\\
&&\Ds ~~~~~~~~~517.5~~~+~~~6.9401~~~+~~~~359.63~~~~-560.63
.~~~~~~~~~~~\nonumber \\
&&\Ds c_4 \approx 11247.97\,. \la{R-c4numbers}
\ea
 \end{subequations}
At the first BLM step, we do not obtain a significant profit in the first coefficient
$c_2 \to c'_2$, as it was in the previous case of $r'_2$.
But the next order coefficient $c_3\approx 352.05$ in (\ref{R-c3numer}) diminished by  two orders
(!)
of magnitude, $c_3 \to c'_3\approx 3.444$
at $a_s(Q^2) \to a'_s=a_s(Q^2 e^{-2}\approx Q^2\cdot 0.135)$,
\begin{subequations}
\label{eq:R1-C}
\ba
&& c'_2  =  -\frac{11}{3};  \la{R1-c2}\\
&&\Ds c'_3 =  \beta_0^2 \cdot2.389~~ - ~~\beta_1 \cdot 1.892~~~ + \beta_0 \cdot 54.63~-560.63\,
\approx \underline{3.444}
 ;\la{R1-c3numer} \\
&&\Ds ~~~~~~~~~~~193.5~~~~-~~~~~121.06~~~+~~~491.63~~-560.63  ~~~~~~~~~~~\nonumber \\
&&\Ds c'_4 \approx 6361.0\,. \la{R1-c4numbers}
\ea
 \end{subequations}
It is interesting that the far fourth coefficient $c_4$, (\ref{R-c4numbers}), reduces twice, $c_4 \to
c'_4$,
(\ref{R1-c4numbers}).
At the second step (PMC) $\bar{\Delta}_1 \approx 7.32$ and
$a^{'}_s \to a^{''}_s=a_s(Q^2\cdot \exp\left[-2-7.32 \beta_0 a'(Q^2)\right])$;
so the region of applicability of PT is shifted far from the scale of a few GeV$^2$.
While the value of $|c''_3|$ goes up to the previous order
of magnitude, compare (\ref{R1-c3numer}) with (\ref{R2-c3numbers}),
\ba
&& c''_2=-\frac{11}{3};~\Ds c''_3\approx -\underline{560.63}\, .\la{R2-c3numbers}
\ea

It is instructive to compare this result with one from seBLM (Sec.~\ref{sec:seBLMPMC}),
where we remove the first 2 terms in (\ref{R1-c3numer}) converting them into
the normalization scale and holding the last two terms in $c''_3$.
For this prescription we obtain
$\Delta_1 \approx 1.25$,
$$a^{'}_s \to a^{''}_s=a_s\left(Q^2 \exp\left[-2-1.25 \beta_0
a'(Q^2)\right]\right)~~~\text{and}~~~c''_3 \approx - 69$$
that looks moderate but is not optimal yet in the sense of series convergence.

Both aforementioned examples demonstrate better convergence at the first BLM step
but fail for the optimization of PT at the second PMC step.
The reason is the different sign of the terms of $r_n~(c_n)$, see the discussion in Sec.6-7 in
\cite{MS04}.
It is clear that one should not remove and  absorb \textit{all the $\beta$-terms} for the
PT-optimization
but leave a part of them for complete cancellation with the $d_n[0]$-term.
We shall treat the circumstances in this way in the next section.

\section{Optimization of the generalized BLM procedure}
 \label{OptBLM}
Indeed, it is not mandatory to absorb all the  $\beta$-terms
as a whole into the new scale $\Delta_1~(\Delta_i)$ following BLM/PMC, but
take instead only those parts of it that are appropriate for
optimization (nullification) of the current order coefficient $r_3~(r_{i+2})$.
At the same time, one should care for the size of the $\Delta_i$ -- PT
coefficients for the shift of scale $\Delta$ in (\ref{Delta}) --
not to violate just this expansion.

Let us consider the optimization of $R^{\rm NS}$ at the second BLM step starting
with the first step expressions in Eqs.(\ref{R1}) and using the general results in
(\ref{1-stage-R3c},\ref{1-stage-R4e},\ref{1-stage-R34d}).
This expression for $r''_3$ can be rewritten as
\ba
&&r''_3=r'_3 -\beta_0^2 \bm{\RedTn{\Delta_{1}}}=
r_3-\beta_0^2d_2[1]^2-\beta_1d_2[1]-\beta_0 2d_2[0]d_2[1]-\beta_0^2 \bm{\RedTn{\Delta_{1}}}
\,.
\ea
The optimization requirement, e.g., $r''_3=0$ leads to the expressions for $\Delta_{1}$ and $r''_4$
\begin{subequations}
\label{eq:opt-r3-4}
 \ba
r''_3=0 \Rightarrow &&\Delta_{1} =r'_3/\beta_0^2= r_3/\beta_0^2 -
d_2[1]^2-\beta_1/\beta_0^2~d_2[1]-1/\beta_0~2d_2[0]d_2[1], \\
&&r''_4=r'_4 -r'_3\left(\beta_1/\beta_0+2d_2[0] \right)\,.
\ea
 \end{subequations}
Numerical calculation at $n_f=4$ gives the estimates for the values of the quantities in
Eqs.(\ref{eq:opt-r3-4}),
\begin{subequations}
\label{eq:opt-r3-4numbers}
 \ba
 &&r''_3=0, ~~\Delta_{1} \approx -3.7, \\
&& r''_4 \approx -4740.52\,, \label{eq:opt-r4numbers} \\
&& a'_{s} \to a^{''}_s=a_s \left(s\cdot e^{-0.692+3.7 \beta_0 a'_s(s) }\right)\,.
\label{eq:opt-R-a''}
\ea
 \end{subequations}
One may conclude that the PT expansion
\ba
R^\text{NS}=1+3C_\text{F}\left\{a''_s+ \frac{1}3 \cdot (a''_s)^2 + 0 \cdot (a''_s)^3+ r''_4 \cdot
(a''_s)^4+\ldots \right\} \label{eq:R3opt }
\ea
significantly improves: \\
(i) $r'_3=0$, while the value of $r''_4$ in
(\ref{eq:opt-r4numbers}) is less than  in (\ref{R-r4numbers}, \ref{R1-r4numbers})
and reduces twice in comparison with the PMC estimate in (\ref{R2-r4numbers}) (taken for $n_f=4$).
\\
(ii) Domain of applicability of the approach extends to a wider region due to the opposite sign at
$\Delta_{1}$,
compare with one for PMC in (\ref{R2-r3numbers}).
This makes the NLO ``shift'' $\Delta$ less, which tends numerically to $0$
 at the boundary of applicability,
 $\Delta=d_2[1]+ \Delta_1 \beta_0 a'_s(s)\approx -0.692+3.7 \beta_0 a'_s(s)$.

\begin{figure}[t]
\includegraphics[width=0.46\textwidth]{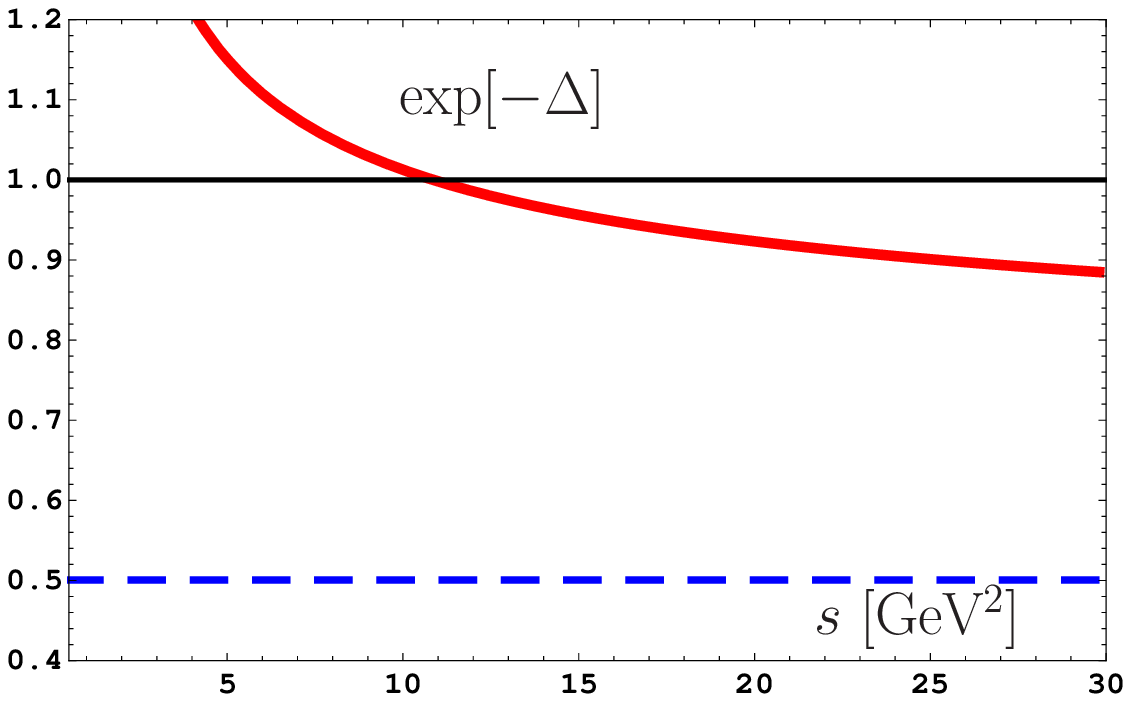}~~
\includegraphics[width=0.47\textwidth]{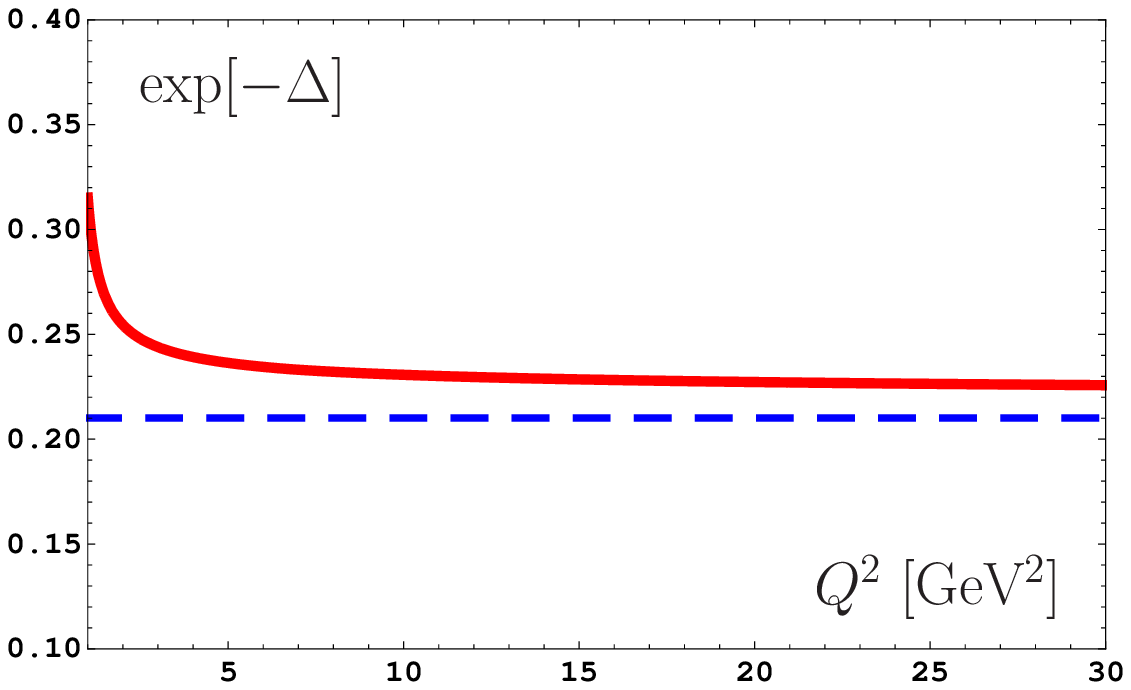}
\caption{\label{fig:NLO-R-factor} \footnotesize
 Factors $\exp\left[-\Delta \right]$ at coupling scale: (Left) for $R^\text{NS}(s)$. Solid (red)
upper line -- the NLO factor
$\exp\left[-0.692+3.7 \beta_0 a'_s(s)\right]$;
long dashed (blue) line -- the LO BLM one $\exp[-0.692]$.
(Right) for $C_{\rm NS}^\text{Bjp}(Q^2)$. Solid (red) upper line -- the NLO factor
$\exp\left[-1.56+ 0.396 \beta_0 a'_s(Q^2)\right]$,
long dashed (blue) line -- the LO  $\exp[-1.56]$.
}
\end{figure}

Indeed, following the usual PT condition $|d_2[1]|\gtrsim |\Delta_1 \beta_0 a'_s(s)|$ or $\Delta
\lesssim 0$
we get for the boundary $s\gtrsim 10$~GeV$^2$, as it is illustrated in
Fig.~\ref{fig:NLO-R-factor}(Left).
The factor $\exp\left[-\Delta \right]$, entering in the argument of $a''_s$
in Eq.(\ref{eq:opt-R-a''}),
see solid (red) upper line in Fig.~\ref{fig:NLO-R-factor}(Left),
satisfies the conditions $1 \gtrsim \exp[-0.692+3.7 \beta_0 a'_s(s)] >1/2$,
this factor slowly decreases with $s$ from the value 1.
It looks tempting to get and use the exact solution for the coupling $a_s^{\star}$,
following from Eq.(\ref{Delta}),
$$a_s^{\star}(s)=a_s\left(s\exp[-\Delta_0 - \Delta_1 \beta_0 a_s^{\star}(s)]\right),$$
rather than its iteration $a''_s(s)$.
It is easy to obtain the useful inequality $a'_s(s) > a_s^{\star}(s) > a''_s(s)$;
moreover, the numerical calculation gives that the difference between $a_s^{\star}$
and $a''_s$ becomes noticeable below $s=1$ GeV$^2$ for this optimized quantity
and for the next one discussed below.

Similar optimization can be performed for $C_\text{NS}^\text{Bjp}(Q^2)$ ($n_f=4$).
We apply  the general combined equations, 
analogous to the ones of Eqs.(\ref{eq:rearrange}).
In these $C_{NS}^{Bjp}$ -oriented expressions 
 we fix  the conditions $c_2^{''}=0$ and  
$c_3^{''}=0$. This leads to the following equations:  
\begin{subequations}
 \label{eq:opt-r3-4numbers}
 \ba
&&c'_2=0,~~\Delta_{0}=c_2/\beta_0 \approx 1.56,,\\
&&a_{s} \to a^{'}_s=a_s \left(Q^2\cdot e^{-1.56}\right)
\ea
 \ba
&&c''_2=c'_2=0, \\
&&c''_3=0,~~\Delta_{1} \approx -0.396, \\
&&c''_4 \approx 4184.64\,, \label{eq:opt-r4numbers} \\
&& a'_{s} \to a^{''}_s=a_s \left(Q^2\cdot e^{-1.56+ 0.396 \beta_0 a'_s(Q^2)} \right)\,.
\label{eq:opt-C-a''}
\ea
 \end{subequations}
The new ``optimized scale'' behaviour of factor $\exp\left[-\Delta \right]$ is illustrated
in Fig.~\ref{fig:NLO-R-factor}(Right) by solid (red) line,
while the broken (blue) line there corresponds to the condition $c'_2=0,~\Delta_{0}=c_2/\beta_0$
that is not the BLM one.
This transformation significantly improves the perturbative series for $C_\text{NS}^\text{Bjp}$,
\ba
\label{eq:C3opt}
C^\text{Bjp}_{\rm NS}(Q^2)=1-3C_\text{F}\left\{a''_s+ 0 \cdot (a''_s)^2 + 0 \cdot (a''_s)^3+
c''_4 \cdot (a''_s)^4+\ldots \right\}
\ea
in comparison with Eqs.(\ref{eq:R-C}, \ref{eq:R1-C}, \ref{R2-c3numbers}).
We conclude that for  both of the considered quantities the PT series are improved,
the corresponding Eqs.(\ref{eq:opt-R-a''},\ref{eq:R3opt }) and
Eqs.(\ref{eq:opt-C-a''},\ref{eq:C3opt})
consist of the main results of this Sec.
We did not perform the next step of optimization with the coefficients $r''_4,~c''_4$ because
in this case we lose control under accuracy.

It is clear that Eqs.(\ref{eq:opt-r3-4numbers},\ref{eq:R3opt }) and
Eqs.(\ref{eq:opt-r3-4numbers},\ref{eq:C3opt})) are not unique optimal solutions because
different efficiency functions may be called ``optimal''.
Therefore, one can satisfy his own efficiency function with the coefficients $\{c_i\}$ basing on
the combined Eqs.(\ref{eq:rearrange}) in the plane $(\Delta_{0},~\Delta_{1})$
or space $(\Delta_{0},~\Delta_{1},~\Delta_{2}, \ldots)$ of fitting free parameters  $\Delta_{j}$.

\section{Conclusion}
\label{sec:concl}
We have considered the general structure of perturbation expansion of renormalization group 
invariant quantities in \text{MS}-schemes to clarify the effects of charge renormalization
and the conformal symmetry breakdown.
Following  the line started in \cite{MS04} we arrived at the matrix representation for this
expansion,
named the $\{\beta\}$-expansion \cite{Kataev:2010du}, instead of the standard perturbation series.
We discussed in great detail the unambiguity of this representation for Adler $D^\text{NS}$-function
(or related $R_{e^+e^-}$--ratio)
and for the Bjorken polarized sum rule $S^\text{Bjp}$ (with the coefficient function
$C^\text{Bjp}_\text{NS}$)
for DIS in order ${\cal O}(\alpha_s^3)$.
The expansion for $S^\text{Bjp}$ was obtained by using different parts of the Crewther relation
\cite{Kataev:2010du} for $D^\text{NS}$  and the coefficient function $C^\text{Bjp}_\text{NS}$.
Others attempts of this presentation \cite{Brodsky:2011ig,Brodsky:2011ta,Mojaza:2012mf,Wu:2013ei}
were discussed too.
We provided  new prediction for $C^{\rm Bjp}_\text{NS}(a_s, n_f, n_{\tilde g})$
with the MSSM massless gluino $n_{\tilde g}$ in order ${\cal O}(\alpha_s^3)$ in
Eqs.(\ref{eq:c1-3beta}),
Sec.\ref{sec:rev-exp.1}, as a byproduct of our consideration.

Based on the $\{\beta\}$-expansion we constructed renormalization group transformation
for the perturbation series of the considered quantities, Eqs.(\ref{eq:rearrange}) in
Sec.\ref{sec:optBLM}.
The initial expansion was split into two parts:
A new series for the expansion coefficients, while the other one -- for the shift of the
normalization scale of the coupling $\alpha_s$.
The contributions from each order can be balanced between these two series.
Different procedures of the PT optimization, including PMC \cite{Brodsky:2011ta,Mojaza:2012mf},
and seBLM \cite{MS04},
were discussed and illustrated by numerical estimates.
We conclude that the corrected  PMC does not provide better PT series convergence
and suggest our own scheme of the series optimization in order of ${\cal O}(\alpha_s^4)$;
the working formulae for $R^\text{NS}$ of the $R_{e^+e^-}$--ratio and $C^{\rm Bjp}_\text{NS}$
are presented in Sec.\ref{OptBLM}.

\acknowledgments
We would like to thank A.~V. Bednyakov, K. G. Chetyrkin,
A.~G. Grozin, V. L. Khandramai, and  N.~G. Stefanis for the fruitful discussion.
The work is done within the scientific program of the
Russian Foundation for Basic Research, Grant No.\ 14-01-00647.
The work by A.K. was supported in part by 
the Russian Science Foundation, Grant N 14-22-00161.
The work of M.S. was supported in part by the BelRFFR--JINR, grant F14D-007.
\begin{appendix}
\appendix
\section{Explicit formulas for $\beta(n_f, n_{\tilde{g}})$ and  $D(n_f, n_{\tilde{g}})$}
 \renewcommand{\theequation}{\thesection.\arabic{equation}}
\label{App:A}   \setcounter{equation}{0}
The required $\beta$-function coefficients with the Minimal Supersymmetric Model (MSSM)\ light
gluinos
\cite{Clavelli:1996pz} calculated in the \MSbar scheme are,
\begin{subequations}
 \label{eq:beta0-3}
\begin{eqnarray}
 \label{eq:beta-b0}
  \beta_0\left(n_f, n_{\tilde{g}}\right)
   &=& \frac{11}{3} C_A - \frac{4}{3}\left( T_R n_f + \frac{n_{\tilde{g}} C_A}{2}\right)\,;\\
 \label{eq:beta-b1}
  \beta_1\left(n_f, n_{\tilde{g}}\right)
     &=& \frac{34}{3}C_A^2
       - \frac{20}{3}C_A \left( T_R n_f + \frac{n_{\tilde{g}} C_A}{2}\right)
      -4\left( T_R n_f C_\text{F}+ \frac{n_{\tilde{g}} C_A}{2} C_A\right);\\
 \label{eq:beta-b2}
  \beta_2\left(n_f, n_{\tilde{g}}\right)
     &=& \frac{2857}{54}C_A^3
       - n_f T_R \left( \frac{1415}{27}C_A^2 +\frac{205}{9}C_A C_F -2 C_F^2 \right)
       + (n_f T_R)^2 \left( \frac{44}{9}C_F +\frac{158}{27}C_A \right) - \nonumber \\
       && \frac{988}{27}n_{\tilde{g}} C_A (C_A^2) +
       n_{\tilde{g}} C_A n_f T_R \left( \frac{22}{9}C_A C_F +\frac{224}{27}C_A^2 \right)
       +(n_{\tilde{g}} C_A)^2 \frac{145}{54} C_A \, .
\end{eqnarray}
 \end{subequations}
The $\beta_3$ coefficient, which includes the MSSM\ light gluinos, is not yet known, so
 we present it here in the standard \cite{RVL97, M.Czakon:2005} simplest form
\begin{eqnarray}
 \beta_3(n_f) &=&
C_A^4 \left( \frac{150653}{486} - \frac{44}{9} \zeta_3 \right)+
C_A^3 T_R n_f \left( - \frac{39143}{81} + \frac{136}{3} \zeta_3 \right)
+ C_F^2 T_R^2 n_f^2 \left( \frac{1352}{27} - \frac{704}{9} \zeta_3
\right) \nonumber \\
 && +C_A C_F T_R^2 n_f^2 \left( \frac{17152}{243} + \frac{448}{9} \zeta_3 \right)
 + C_A C_F^2 T_R n_f \left( - \frac{4204}{27} + \frac{352}{9} \zeta_3 \right) + \frac{424}{243} C_A
 T_R^3 n_f^3 \nonumber \\
&&+ C_A^2 C_F T_R n_f \left( \frac{7073}{243} - \frac{656}{9} \zeta_3 \right)
+ C_A^2 T_R^2 n_f^2 \left( \frac{7930}{81} + \frac{224}{9} \zeta_3
\right) + \frac{1232}{243} C_F T_R^3 n_f^3 \nonumber \\
&&+ 46 C_F^3 T_R n_f + n_f \dRANA \left( \frac{512}{9} - \frac{1664}{3} \zeta_3 \right) + n_f^2
\dRRNA \left( - \frac{704}{9} + \frac{512}{3} \zeta_3
\right) \nonumber \\
&&+ \dAANA \left( - \frac{80}{9} + \frac{704}{3} \zeta_3 \right) \label{eq:beta3}.
\end{eqnarray}
For the $SU_{c}(N)$-group with fundamental fermions the invariants read:
\begin{equation}
T_R = \frac{1}{2}, \; C_F = \frac{N^2-1}{2N}, \; C_A =N,\;
d^{abc}d^{abc}= \frac{(N^2-4)N_A}{N};~N_A = 2C_\text{F} C_\text{A} \equiv N^2-1.
\end{equation}
\begin{equation}
\dRANA = \frac{N(N^2+6)}{48}, \;\;\;\; \dAANA =
\frac{N^2(N^2+36)}{24}, \dRRNA = \frac{N^4-6N^2+18}{96N^2}.
\end{equation}

The $D^{\rm NS}$-function evaluated  in \cite{Chetyrkin:1996ez} in the same
model in the case when the masses of gluion are neglected\footnote{In the numerical
case this expression from  \cite{Chetyrkin:1996ez} coincides
with the result of the related numerical calculation of
\cite{Kataev:1983at}.}
reads
\begin{subequations}
 \label{eq:D1-3}
  \begin{eqnarray}
&& D^{\rm NS}(a_s, n_f, n_{\tilde g})=
               1+  a_s\cdot(3C_F)  \label{eq:D-1}
 \end{eqnarray}
 \begin{eqnarray}
  &&     +a_s^2
          \left\{-\frac{3}{2}C_F^2
             +2C_F\left[\frac{123}{2}-44\zeta_3
                           -(11-8\zeta_3)n_{\tilde g} \right]\frac{C_A}2
                             -2C_F(11-8\zeta_3)n_fT_R\right\}
                                             \label{eq:D-2}           \\
  &&    + a_s^3
          \left\{
         -\frac{69}{2} C_F^3
  -C_F^2C_A\left[127+572\zeta_3-880\zeta_5
      -(36+104\zeta_3-160\zeta_5)n_{\tilde g} \right]
                                                        \right. \nonumber\\
  &&
      \left.     +C_F C_A^2\left[\frac{90445}{54}-\frac{10948}{9}\zeta_3
                        -\frac{440}{3}\zeta_5
       -\left(\frac{33767}{54}-\frac{4016}{9}\zeta_3
                        -\frac{80}{3}\zeta_5 \right)n_{\tilde g} \right. \right.
                                                                \nonumber\\
  &&
      \left. \left. +\left(\frac{1208}{27}-\frac{304}{9}\zeta_3 \right)
                                                   n_{\tilde g}^2
                                             \right]
          -n_fT_RC_F^2[29-304\zeta_3+320\zeta_5]      \right.    \nonumber\\
  &&
       \left.     -n_fT_RC_FC_A\biggl[\frac{31040}{27}-\frac{7168}{9}\zeta_3
                                 -\frac{160}{3}\zeta_5
           -\biggl( \frac{4832}{27}-\frac{1216}{9}\zeta_3 \biggr)
                                                 n_{\tilde g}
                                                       \biggr]  \right.  \nonumber\\
  &&  \left.
           +3C_F\left[\frac{302}{9}
                                 -\frac{76}{3}\zeta_3\right]\left(\frac{4}3 T_Rn_f\right)^2
             \right\}= \label{eq:D-3}
\end{eqnarray}
 \end{subequations}
\begin{subequations}
 \label{eq:D1-3mod}
 \begin{eqnarray}
&& =1+  a_s(3C_F) + a_s^2(3C_F)\cdot
          \left\{\frac{\rm C_A}3-\frac{\rm C_F}2+
          \left(\frac{11}2-4\zeta_3\right)\beta_0\left(n_f, n_{\tilde{g}}\right) \right\}\\
          \label{D3:gl}
&&+a_s^3(3C_F)\cdot
          \left\{ \left(\frac{302}9-\frac{76}3\zeta_3\right)\beta_0^2\left(n_f,
          n_{\tilde{g}}\right)+
          \left(\frac{101}{12}-8\zeta_3\right)\beta_1\left(n_f, n_{\tilde{g}}\right)\right. \nonumber
          \\
&&        \left.  +\left[{\rm C_A}\left(\frac{3}4 + \frac{80}3\zeta_3 -\frac{40}3\zeta_5\right) -
    {\rm C_F}\left(18 + 52\zeta_3 - 80\zeta_5\right)\right]\beta_0\left(n_f,
    n_{\tilde{g}}\right)\right. \nonumber \\
&&     \left. + \left(\frac{523}{36}-
        72 \zeta_3\right){\rm C_A^2}
    +\frac{71}3 {\rm C_A C_F} - \frac{23}{2} {\rm C_F^2}       \right\}
\end{eqnarray}
  \end{subequations}
  The Bjorken coefficient function $C^\text{Bjp}_\text{NS}$ of the DIS sum rules calculated first in
  \cite{Larin:1991tj}
\begin{subequations}
 \label{eq:C1-3}
  \begin{eqnarray}
 C^\text{Bjp}_\text{NS}(a_s, n_f)=
               1 + &a_s&(-3C_F)  \label{eq:C-1} \\
        +&a_s^2&(-3C_F)
          \left[-\frac{7}{2}C_F
             + \frac{23}3C_A
                             -\frac{8}{3} T_Rn_f\right] \label{eq:C-2}
 \end{eqnarray}
 \begin{eqnarray}
  &&    + a_s^3(-3C_F)
          \left\{
         \frac{C_F^2}{2}
  +C_FC_A\left[\frac{176}{9}\zeta_3- \frac{1241}{27} \right]
  + C_A^2\left[\frac{10874}{81} -\frac{440}{9}\zeta_5 \right]  \right.\\
   &&   \left.           +n_fT_R C_F\left[\frac{133}{27}-\frac{80}{9}\zeta_3\right]
   -n_fT_R C_A\biggl[\frac{7070}{81}+16\zeta_3
                                 -\frac{160}{9}\zeta_5\biggr]  \right.  \nonumber\\
  &&  \left.
            + \frac{115}{18}\left(\frac{4}3n_fT_R\right)^2
             \right\}~. \label{eq:C-3}
\end{eqnarray}
 \end{subequations}
 The prediction for $C^\text{Bjp}$ obtained in Sec.\ref{sec:rev-exp.1} of this article under the same
 conditions
 as Eq.(\ref{eq:D1-3}) reads
\begin{subequations}
 \label{eq:C1-3mod}
 \begin{eqnarray}
&&C^\text{Bjp}_\text{NS}(a_s, n_f, n_{\tilde g}) =1+  a_s(-3C_F)  \\
&&+a_s^2(-3C_F)\cdot
          \left\{\frac{1}{3}{\rm C_A}-\frac{7}{2}{\rm C_F}+
          2\beta_0\left(n_f, n_{\tilde{g}}\right) \right\}\\ \label{D3:gl}
&&+a_s^3(-3C_F)\cdot
          \left\{\frac{115}{18}\beta_0^2\left(n_f, n_{\tilde{g}}\right)+
          \left(\frac{59}{12}-4\zeta_3\right)\beta_1\left(n_f, n_{\tilde{g}}\right)\right. \nonumber
          \\
&&        \left.  -\left[\left(
\frac{215}{36}- 32 \zeta_3+\frac{40}{3}\zeta_5\right){\rm C_A}
    +\bigg(\frac{166}{9}- \frac{16}3\zeta_3\bigg){\rm C_F}\right]\beta_0\left(n_f,
    n_{\tilde{g}}\right)\right. \nonumber \\
&&     \left. +
 \bigg( \frac{523}{36} - 72\zeta_3\bigg){\rm C_A^2}+\frac{65}{3}{\rm C_F C_A}+ \frac{\rm C_F^2}{2}
 \right\}.
\end{eqnarray}
  \end{subequations}
\section{Natural forms for $\beta$-expansion of $D^\text{NS}$ and $C^\text{NS}$ }
 \renewcommand{\theequation}{\thesection.\arabic{equation}}
  \label{App-B}\setcounter{equation}{0}
 Here we present for completeness the results of (\ref{eq:d1-4},\ref{eq:c1-4}) in their 
 ``natural form'' changing only the normalization factors \cite{Kataev:2010du}, 
 which correspond to the coupling $\Ds \frac{\alpha_s}{\pi}$
with $\Ds \beta_0=\frac{1}4 \left( \frac{11}3 C_\text{A} - \frac{4}{3}\left(T_R n_f + n_{\tilde{g}}
C_A\frac{1}{2}\right)\right), \ldots$,

\begin{subequations}
\label{eq:Bd1-4 }
\begin{eqnarray}
&&d^\text{NS}_1=\frac{3}{4}{\rm C_F}, \\
&&~d^\text{NS}_2[1]= \bigg(\frac{33}{8} - 3\zeta_3\bigg){\rm C_F},~
d^\text{NS}_2[0]= -\frac{3}{32}{\rm C_F^2}  + \frac{1}{16}{\rm C_FC_A},
\label{eq:BD-2}
\end{eqnarray}
 \begin{eqnarray}
 &&d^\text{NS}_3[2]=\bigg(\frac{151}{6}-19\zeta_3\bigg){\rm C_F},~
 d^\text{NS}_3[0,1]= \bigg(\frac{101}{16}-6\zeta_3\bigg){\rm C_F}, \label{eq:BD-32}\\
&& d^\text{NS}_3[1]=
    \bigg(-\frac{27}{8} - \frac{39}{4}
    \zeta_3 + 15\zeta_5\bigg){\rm C_F^2}
-\bigg(\frac{9}{64} - 5\zeta_3 +\frac{5}{2}\zeta_5\bigg)
{\rm C_FC_A} \label{eq:BD-31},\\
 &&
~d_3[0]= - \frac{69}{128}{\rm  C_F^3}+ \frac{71}{64}{\rm C_F^2C_A}+
    \bigg(\frac{523}{768}- \frac{27}{8}\zeta_3\bigg){\rm C_FC_A^2}~. \label{eq:BD-30}
\end{eqnarray}
 \end{subequations}
\begin{subequations}
\label{eq:Bc1-4 }
\begin{eqnarray}
c^\text{NS}_1&=&-\frac{3}{4}{\rm  C_F},~\\
c^\text{NS}_2[1]&=& -\frac{3}{2}{\rm C_F},~
c^\text{NS}_2[0]= \frac{21}{32}{\rm C_F^2}-\frac{1}{16}{\rm C_FC_A},
\end{eqnarray}
\begin{eqnarray}
c^\text{NS}_3[2]&=&-\frac{115}{24}{\rm C_F},~
c^\text{NS}_3[1]= \bigg(\frac{83}{24}- \zeta_3\bigg){\rm C_F^2} +
\bigg(\frac{215}{192}- 6 \zeta_3+\frac{5}{2}\zeta_5\bigg){\rm C_FC_A},
\label{eq:Bc-31} \\
c^\text{NS}_3[0,1]&=&\bigg(-\frac{59}{16}+3\zeta_3\bigg){\rm C_F},~ \\
c^\text{NS}_3[0] &=& - \frac{3}{128} {\rm C_F^3} -\frac{65}{64} {\rm C_F^2C_A}-
\bigg(   \frac{523}{768} - \frac{27}{8}\zeta_3\bigg){\rm C_FC_A^2}. \label{eq:Bc-30}
\end{eqnarray}
 \end{subequations}

\section{R-ratio}
 \renewcommand{\theequation}{\thesection.\arabic{equation}}
   \label{App-D}\setcounter{equation}{0}
\begin{table}[th]
\caption{The table of the $T^{mk}$-matrix.
The one-loop contributions are marked by black, two-loop contributions are
 marked by red,
 three-loop contribution --- by blue, while the four-loop contribution is colored by green.
\label{Tab:r_n.d_k}}
\begin{tabular}{|c||p{24mm}|p{24mm}|p{24mm}|p{24mm}|p{24mm}|p{24mm}|}\hline \hline
                     &\centerline{$\bm{d_1}$}
                                   &\centerline{$\bm{d_2}\vphantom{^\Big|}$}
                                                &\centerline{$\bm{d_3}$}
                                                             &\centerline{$\bm{d_4}$}
                                                                          &\centerline{$\bm{d_5}$}
                                                                                    &\centerline{$\bm{d_6}$}
\\ \hline \hline
          $\bm{r_1}\vphantom{^{\Big|}}$
                     & \centerline{$\bm{1}$}
                                   &            &            &            &          &
\\ \hline
          $\bm{r_2}\vphantom{^{\Big|}}$
                     & \centerline{$0$}
                                   & \centerline{$\bm{1}$}
                                                &            &            &          &
\\ \hline
          $\bm{r_3}\vphantom{^{\Big|}}$
                     &
                     \centerline{$\Ds-\frac{(\pi\,\beta_0\vphantom{^{\big|}})^2}{3}\vphantom{^{\big|}_{\big|}}$}
                                   & \centerline{$0$}
                                                & \centerline{$\bm{1}$}
                                                             &            &          &
\\ \hline
          $\bm{r_4}\vphantom{^{\Big|}}$
                     &\centerline{$0$}
                      \centerline{\RedTn{$\Ds\bm{-\frac{5\,\pi^2\vphantom{^{\big|}}}{6}\,\beta_0\,\beta_1}\vphantom{^{\big|}_{\big|}}$}}
                                   & \centerline{$\Ds-\frac{(\pi
                                   \beta_0\vphantom{^{\big|}})^2}{3}~3\vphantom{^{\big|}_{\big|}}$}
                                                & \centerline{0}
                                                             &\centerline{$\bm{1}$}
                                                                          &            &
\\ \hline
          $\bm{r_5}\vphantom{^{\Big|}}$\footnote{\footnotesize This expression for $r_5$ was
          presented first in \cite{Kataev:1995vh}}
                     &\centerline{$\Ds\frac{(\pi\,\beta_0\vphantom{^{\big|}})^4}{5}\vphantom{^{\big|}}$}\
                      \centerline{\RedTn{\bm{$\Ds -\frac{\pi^2}{2}\,\beta_1^2}$}}
                      \centerline{\BluTn{$\Ds
                      \bm{-\pi^2\,\beta_0\,\beta_2}\vphantom{^{\big|}_{\big|}}$}}
                                   &\centerline{$\Ds0\vphantom{\frac{(\pi\,\beta_0^{\big|})^4}{5}_{\big|}}$}
                                    \centerline{\RedTn{$\Ds
                                    \bm{-\frac{7\,\pi^2}{3}\,\beta_0\,\beta_1}\vphantom{^{\big|}}$}}
                                                &\centerline{$\Ds-\frac{(\pi \beta_0)^2}{3}\,6$}
                                                             &\centerline{0}
                                                                          &\centerline{$\bm{1}$}&
\\ \hline
   $\bm{r_6}\vphantom{^{\Big|}}$
                     &\centerline{$0$}
                      \centerline{\RedTn{$\Ds\bm{-\frac{77\,\pi^4\vphantom{^{\big|}}}{60}\,\beta_0^3\,\beta_1}
                      \vphantom{^{\big|}_{\big|}}$}}
                      \centerline{\BluTn{$\Ds
                      \bm{-\frac{7\pi^2}{6}\,\beta_1\,\beta_2}\vphantom{^{\big|}_{\big|}}$}}
                      \centerline{\DarkGreen $\Ds \bm{ -\frac{4\,\pi^2}{3}\,\beta_0\,\beta_3}
                      \vphantom{^{\big|}_{\big|}}$}
                                   & \centerline{$\Ds-\frac{(\pi
                                   \beta_0\vphantom{^{\big|}})^4}{5}~5\vphantom{^{\big|}_{\big|}}$}
                                     \centerline{$\Ds\RedTn{\bm{
                                     -\frac{4\,\pi^2}{3}\,\beta_1^2~}}\vphantom{^{\big|}_{\big|}}$}
                                     \centerline{$ \Ds  \BluTn{\bm{
                                     -\frac{8\,\pi^2}{3}\,\beta_0\,\beta_2}}$}
                                                & \centerline{0} 
                                           \centerline{$ \Ds
                                           \RedTn{\bm{-\frac{9\,\pi^2}{2}\,\beta_0\,\beta_1}}\vphantom{^{\big|}_{\big|}}$}
                                                             &
                                                        $\Ds -\frac{(\pi\,\beta_0)^2}{3}\,10$    &
                                                        \centerline{$0$}
                                                        &\centerline{$\bm{1}$}
\\  \hline\hline
\end{tabular}
 \end{table}
Table~\ref{Tab:r_n.d_k} exemplifies the structure of a few first
 coefficients $r_m$ of the conventional expansion of the $R$-ratio.
 Every coefficient $r_m$ contains a number of $d_{k}~(k\leq m)$ terms
 in its expansion, which are shown in the corresponding row.
 In other words, $r_m=T^{mk}d_k$ (summation in $k=1,\ldots,m$ is assumed),
 where $T^{mk}$ are the Table entries.
  Note that the content of this table is limited here only by the restricted place.
\section{Explicit formulae for $\Delta_{i}$ }
 \renewcommand{\theequation}{\thesection.\arabic{equation}}
   \label{App-E}\setcounter{equation}{0}
The explicit expressions for the elements of the proper scales
$\Delta_1$ and $\Delta_2$ are given by
 \ba
&&\Delta_{0} = d_2[1];\\
&&\Delta_{1} = d_3[2]- d^2_2[1]-\BluTn{\pi^2/3 }+\frac{\beta_1}{\beta_0^2}\left(d_3[0,1] -
d_2[1]\right)
+\frac{1}{\beta_0} \left( \underline{\underline{d_3[1]-2d_2[1]d_2[0]}}\right); \\
&&\Delta_{2} =
\left(d_4[3]-3d_2[1]d_3[2]+2(d_2[1])^3 \BluTn{-\pi^2 d_2[1]}\right)
+\beta_2/\beta_0^3 (d_4[0,0,1]-d_2[1])+
 \nonumber \\
&&\phantom{\Delta_{0} = } \beta_1/\beta_0^2
\left[d_4[1,1]-3d_3[0,1]d_2[1]+\frac3{2}(d_2[1])^2 - d_3[2] -\BluTn{\pi^2/2}\right]+
\beta_1^2/\beta_0^4 \left(d_2[1]- d_3[0,1]\right) + \nonumber \\
&& \phantom{\Delta_{0} = }
\beta_1/\beta_0^3\left(\underline{\underline{d_4[0,1]-d_3[1]-2d_2[0]\left(d_3[0,1]-d_2[1]\right)}}
\right)+  \\
&& \phantom{\Delta_{0} = }1/\beta_0\left[\underline{\underline{d_4[2]-3d_3[1]d_2[1]+d_2[0]
\left(5d_2[1]^2-2d_3[2]-\BluTn{\pi^2/3} \right)}} \right]+  \\
&& \phantom{\Delta_{0} = }1/\beta_0^2
\left[
\underline{\underline{\underline{d_4[1]-3d_3[0]d_2[1]+2d_2[0](2d_2[1]d_2[0]-d_3[1])}}}\right]\,.
\ea

\end{appendix}
\newcommand{\noopsort}[1]{} \newcommand{\printfirst}[2]{#1}
  \newcommand{\singleletter}[1]{#1} \newcommand{\switchargs}[2]{#2#1}

\end{document}